\documentclass{osa-article}

\journal{oe}

\newcommand{\vect}[1]{\mathbf{#1}}
\newcommand{\vectrho}{\boldsymbol{\rho}}

\newcommand{\secref}[1]{Sec.~\ref{#1}}

\newcommand{\appenref}[1]{Appendix~#1}
\newcommand{\eqnref}[1]{Eq.~\eqref{#1}}

\newcommand{\figref}[1]{Fig.~\ref{#1}}

\newcommand{\citeasnoun}[1]{Ref.~\citenum{#1}}

\newcommand{\newtext}[1]{{#1}}
\renewcommand{\Re}{\operatorname{Re}}

\newcommand{\maxwell}[1]{\nabla\times\frac{1}{\mu_0\mu_r}\nabla\times\vect{E}_{#1}  -\omega_{#1}^2\epsilon_0\epsilon_r(\rho)\vect{E}_{#1}=\ -i\omega_{#1}\vect{J}_{#1}}
\newcommand{\maxwellH}{\left[-\nabla\cdot\frac{1}{\varepsilon(\rho)}\nabla - k^2 \mu_0\mu_r\right] H_z(\boldsymbol{x}) = f(\boldsymbol{x}) }

\newcommand{\TopOpt}{TO}
\newcommand{\code}[1]{\texttt{#1}}

\renewcommand{\d}{\mathop{}\!\mathrm{d}}

\usepackage{lineno}
\usepackage{nicefrac}
\usepackage{comment}
\usepackage{xcolor}
\usepackage{fancyvrb}
\usepackage{fvextra}

\InputIfFileExists{minted-cache/default-pyg-prefix.pygstyle}{}{}
\InputIfFileExists{minted-cache/default.pygstyle}{}{}
\makeatother

\newcommand{\DOF}{\rho}
\newcommand{\DOFfiltered}{\tilde{\rho}}
\newcommand{\DOFsmooth}{\hat{\tilde{\rho}}}
\newcommand{\DOFprojected}{\hat{\rho}}
\newcommand{\Rfilt}{\tilde{R}}
\newcommand{\Rsmooth}{\hat{R}}
\newcommand{\kfilt}{\tilde{k}}
\newcommand{\ksmooth}{\hat{k}}

\begin{document}

\title{Unifying and accelerating level-set and density-based topology optimization by subpixel-smoothed projection}

\author{Alec~M.~Hammond,\authormark{1*} Ardavan~Oskooi,\authormark{1}
Ian~M.~Hammond,\authormark{2} Mo~Chen,\authormark{3} Stephen~E.~Ralph,\authormark{4} and Steven~G.~Johnson\authormark{3}}
\address{$^1$Meta, 1 Hacker Way, Menlo Park, CA 94025, USA \\
$^2$Department of Electrical Engineering and Computer Science, Massachusetts Institute of Technology, Cambridge, MA 02139, USA\\
$^3$Department of Mathematics, Massachusetts Institute of Technology, Cambridge, MA 02139, USA\\
$^4$School of Electrical and Computer Engineering, Georgia Institute of Technology, Atlanta, GA 30308, USA \\}
\email{\authormark{*}alec.m.hammond@gmail.com}


\begin{abstract}
We introduce a new ``subpixel-smoothed projection'' (SSP) formulation for differentiable binarization in topology optimization (TopOpt) as a drop-in replacement for previous projection schemes, which suffer from \newtext{near}-non-differentiability and slow convergence as binarization improves.  Our new algorithm overcomes these limitations by depending on both the underlying filtered design field and its spatial gradient, instead of the filtered design field alone. We can now smoothly transition between density-based TopOpt (in which topology can easily change during optimization) and a level-set method (in which shapes evolve in an almost-everywhere binarized structure).  We demonstrate the effectiveness of our method on several photonics inverse-design problems and for a variety of computational methods (finite-difference, Fourier-modal, and finite-element methods).   SSP exhibits both faster convergence and greater simplicity.
\end{abstract}

\section{Introduction}
\label{sec:intro}

Photonics topology optimization (TopOpt) is a form of ``inverse design''~\cite{Molesky2018} in which freeform arrangements of materials are evolved to maximize a given design objective (e.g.~transmitted power), with millions of degrees of freedom (DOFs) that can be used to explore nearly arbitrary geometries, including arbitrary topologies (e.g.~how many ``holes'' are present).  TopOpt algorithms are divided into an unfortunate dichotomy, however: there are level-set set methods~\cite{van2013level} that can exactly describe discontinuous ``binary'' materials (e.g.~silicon in some regions and air in others) but make it more difficult for optimization to change the topology~\cite{to_approaches}; versus density-based methods~\cite{to_approaches,Jensen2010} that allow the topology to change by smoothly evolving one material to another, but which become more and more difficult to optimize as the structure is ``binarized'' to be everywhere one material or another.
The key difficulty centers around the computation of derivatives: in order to optimize efficiently over many parameters, the design objective must be differentiable with respect to the DOFs, and the derivatives must be well behaved, e.g.~lacking sharp ``kinks'' where the second derivative (Hessian matrix) is suddenly huge in some directions (such an ``ill-conditioned'' Hessian slows optimization convergence~\cite{nocedal2006numerical,bertsekas2016optimization}).  In density-based methods, binarization involves a ``projection'' step in which the material is projected towards one extreme or an other, controlled by a steepness hyper-parameter $\beta$, but the projection becomes more ill-conditioned as $\beta$ increases and the structure becomes more binary (more manufacturable).  In this paper, we develop a new form of projection that allows us to bridge the gap between density-based and level-set methods: we employ density DOFs $\DOF$ that can smoothly describe changes in topology, but our new projection scheme also allows us to increase $\beta$ to $\infty$ (equivalent to a level set) while retaining \newtext{mostly} well-behaved derivatives.

\begin{figure}
    \centering
    \includegraphics{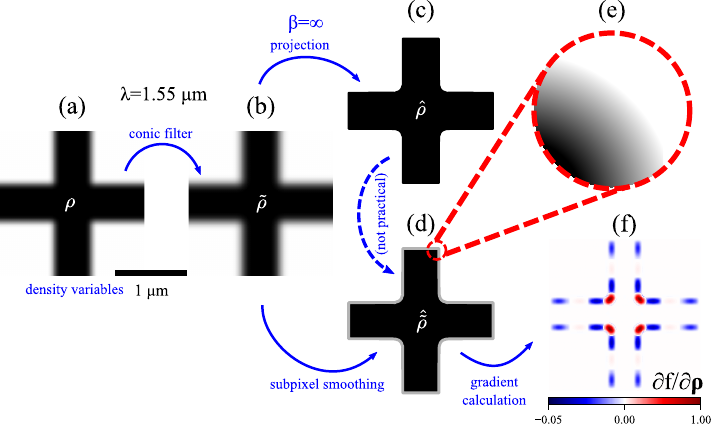}
    \caption{Schematic of ``three-field'' approach to density-based topology optimization: (a)~an initial density field $\DOF$ (here for the waveguide-crossing problem of \figref{fig:waveguide_TO}) is (b)~low-pass filtered to a smoothed/grayscale density $\DOFfiltered$ and then (c)~projected to a binarized density $\DOFprojected$ with steepness hyper-parameter $\beta$, which represents the physical material arrangement.   Step~(c) is not differentiable for a step-function ($\beta = \infty$) projection, but could theoretically be made differentiable by smoothing the edges at infinite spatial resolution to obtain~(d)~$\DOFsmooth$, indicated by the dashed ``not practical'' arrow.  Our new algorithm obtains (d)~$\DOFsmooth$ directly from (b)~$\DOFfiltered$.  The result~$\DOFsmooth$ is almost-everywhere binary except for a thin ``subpixel'' boundary layer~(e) within the grid resolution~$\sim \Delta x$ of the interface. (f)~This layer allows \newtext{an objective function $f$ (here from \secref{sec:crossing})} to have a finite gradient (nonzero only near boundaries) with respect to the density~$\DOF$~(a); \newtext{here, the positive gradients (red) are $\approx 20\times$ larger than the negative gradients (blue).}}
    \label{fig:overview}
\end{figure}

\begin{figure}
    \centering
    \includegraphics{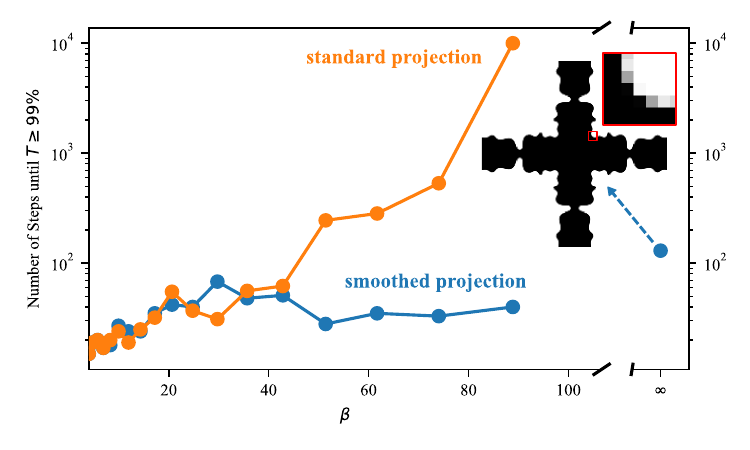}
    \caption{
    \newtext{The number of CCSA~\cite{Svanberg2002} optimization steps required to achieve 99\% transmission for a waveguide-crossing example (\secref{sec:convergence}), using standard projection (orange) and our SSP method (blue) as a function of $\beta$. As expected, standard projection requires a diverging number of optimization iterations to achieve the specified tolerance as $\beta$ increases. In contrast, SSP converges to a finite number of optimization steps as $\beta\to\infty$. Inset: The optimized structure for $\beta=\infty$, along with an enlarged section of the structure illustrating the 1-pixel-thick smoothed interface.
    }
    \label{fig:diverging_optimization}
    }
\end{figure}

In a level-set method, the DOFs are a level-set function $\phi(\vec{x})$ which denotes one material (e.g.~silicon) when $\phi < 0$ and another material (e.g.~air) when $\phi > 0$; unfortunately, when the topology changes (e.g.~two boundaries intersect or a new hole appears), differentiation with respect to $\phi$ requires complicated ``topological derivative'' algorithms~\cite{to_approaches} that can be onerous to implement, so in practice a level-set implementation may be restricted to the initially chosen topology (which is also known as ``shape optimization'' with a fixed topology~\cite{to_approaches}).  In density-based TopOpt (reviewed in \secref{sec:background}), the DOFs are a density function $\DOF(\vec{x})$ for which $\DOF = 0$ represents one material, $\DOF = 1$ represents another material, and intermediate $0 < \DOF < 1$ values represent impractical ``interpolated'' materials (e.g.~interpolating the refractive index)~\cite{meep_adjoint}.  This makes the physical solution easy to differentiate with respect to $\DOF$ and allows the topology to continuously change (e.g.~a new ``hole'' can appear as $\DOF$ goes smoothly from 1 to 0 in some region), but optimization may lead to non-manufacturable designs in which many intermediate materials are present.  To combat this, and also to regularize the problem by introducing a minimum lengthscale, density-based methods introduce two additional density fields~\cite{projection_overview,linus_morph,qian_actuators,zhou_geometric,svanberg_harmonic} (\figref{fig:overview}a--c): a smoothed $\DOFfiltered$ given by the convolution of $\DOF$ with a minimum-lengthscale low-pass filter, followed by a projection $\DOFprojected$ that passes $\DOFfiltered$ through a step function to binarize it to either $0$ or $1$.  A step function is not differentiable, however, so in practice the projection step uses a smoothed ``sigmoid'' function characterized by a steepness parameter $\beta$.  As $\beta$ increases towards $\infty$, the structure becomes more binarized, at the cost of slower optimization convergence \newtext{(\figref{fig:diverging_optimization} and \secref{sec:convergence})} because the derivative approaches zero everywhere except at the step where it diverges (and its second derivative becomes ill-conditioned; \newtext{some authors advocate explicitly limiting step sizes in optimization algorithms for large $\beta$~\cite{Guest2011,Dunning2025}}).   Managing this tradeoff requires cumbersome tuning of the rate at which the hyper-parameter $\beta$ is increased during optimization \newtext{(and even automating this~\cite{Dunning2025} requires additional hyper-parameters)}, in order to obtain a mostly binarized structure with good performance in a reasonable time.  \newtext{Moreover, at any finite $\beta$, it is still possible to obtain arbitrary large regions of intermediate materials by $\DOFfiltered$ values close to the threshold~\cite{van2013level}, so additional penalties or constraints (with their own hyper-parameters) are sometimes necessary to ensure binarization~\cite{jensen2005damping, hansen2024inverse, chen2025}.  While many practical results have been obtained using these methods, it would be much more convenient if full binarization could be ensured without slowing convergence and without additional penalties or complexity.}

To \newtext{attain this goal}, we present a new density projection ($\DOFsmooth$), ``subpixel-smoothed projection'' (SSP), which supports a seamless transition from density-based to level-set optimization. Our approach offers a drop-in replacement for existing projection functions, and allows designers to leverage all their existing topology-optimization software---the only implementation difference is that our~$\DOFsmooth(\vect{x})$ depends on \emph{both}~$\DOFfiltered$ and~$\Vert \nabla\DOFfiltered \Vert$ at~$\vect{x}$, rather than just on~$\DOFfiltered$ as for the old~$\DOFprojected$.   Let the computational mesh spacing/resolution be of order~$\Delta x$.  Conceptually, SSP works \emph{as if} the filtered density $\DOFfiltered$ were first interpolated and projected to yield the conventional $\DOFprojected$ at ``infinite'' resolution (regardless of $\Delta x$), which is \emph{then} convolved with a $\Delta x$-diameter smoothing filter to yield a density $\DOFsmooth$ (depicted in \figref{fig:overview}d--e) that is binary \emph{almost everywhere} (for $\beta=\infty$) as the mesh resolution increases ($\Delta x \to 0$), but which changes \newtext{mostly} differentiably as $\DOF$ changes \newtext{(even for $\beta=\infty$)}.  \newtext{(We use ``almost everywhere'' in the technical sense~\cite{Rudin_1976}: as $\Delta x \to 0$, the structure becomes binary except on a set of measure zero, or within $\sim \Delta x$ of interfaces.  The low-pass filtering, which is set by the minimum desired lengthscale $\Rfilt$ independent of $\Delta x$, ensures that these interface regions approach a set of measure zero.)} As depicted in \figref{fig:overview}, we make this conceptual algorithm \emph{practical} by instead computing $\DOFsmooth$ directly from the filtered density $\DOFfiltered$, with the ``infinite-resolution smoothing'' performed analytically via a planar/linear approximation of the $\DOFfiltered = 0.5$ level set.  In particular, \secref{sec:smoothed_projection} proposes an efficient algorithm, compatible with any dimensionality, that yields a mostly twice-differentiable \newtext{(for any~$\beta$)} pointwise function $\DOFsmooth(\DOFfiltered, \Vert \nabla \DOFfiltered \Vert )$, except at changes in topology where it reduces to~$\DOFprojected$ (\secref{sec:topology_changes}). We demonstrate the flexibility of SSP by applying it in \secref{sec:experiments} to various TopOpt problems with three different Maxwell solvers: finite-difference time domain (FDTD)~\cite{taflove_em_book}, a Fourier modal method (FMM)~\cite{kim2012fourier}, and a finite-element method (FEM)~\cite{jin2015finite}; the same approach is also directly applicable to non-photonics TopOpt problems (\secref{sec:conclusion}). Several test problems are taken from our recently published photonics-testbed suite~\cite{testbed}, and SSP results in final designs comparable to traditional TopOpt, but with faster convergence at large~$\beta$ \newtext{(\figref{fig:diverging_optimization})} and allowing more rapid $\beta$ increases. We also show how SSP can be applied to both topology optimization (by incrementing the projection steepness $\beta$ in a few steps to $\beta=\infty$, \newtext{where a final optimization can now be performed}) as well as to shape optimization (in which the binary design is optimized for a fixed topology \newtext{only at} $\beta=\infty$, which is now differentiable for a fixed topology). We demonstrate that SSP's better-conditioned formulation of the projection problem requires fewer optimization iterations at large $\beta$ than the conventional density-projection scheme \newtext{(\secref{sec:convergence})}. By eliminating the density/level-set dichotomy, we believe that SSP should not only improve performance and simplicity---$\beta$ can be increased to $\infty$ as soon as the topology stabilizes, without sacrificing convergence rate---but may additionally lead to improvements in other aspects of TopOpt, such as techniques for imposing lengthscale constraints or more accurate discretization schemes (\secref{sec:conclusion}).

\newtext{A close relative of SSP can be found in level-set methods with subpixel fill-fraction averaging~\cite{van2013level}.  In these methods, the level-set function $\phi$ (replacing our $\DOFfiltered$) is passed through a Heaviside projection, but the result is discretized onto a fixed mesh by introducing artificial intermediate materials (much-like density-based {\TopOpt}) in mesh elements that intersect the $\phi = 0$ level set, assigned according to the fill-fraction of the $\phi > 0$ region in element.  Typically, this fill fraction is computed analytically by linear interpolation of $\phi$~\cite{deRuiter2004,van2012explicit}, analogous to our linearization of $\DOFfiltered$ below.  For a uniform mesh, we point out in \secref{sec:smoothed_projection} that such a scheme is analogous to SSP with a different smoothing kernel, given by the indicator function of an element.  However, this has three drawbacks compared to SSP: it requires a complex mesh-dependent implementation (e.g.~for a cubic voxel there are many different possible intersections with a plane~\cite{rezk2005vertex}), it is not guaranteed to yield continuous derivatives (a derivative discontinuity occurs in the fill fraction when the level set crosses an element face that it is parallel to, with arbitrarily large second derivatives for near-parallel cases), and smoothly transitioning to/from density-{\TopOpt}'s finite-$\beta$ (sigmoid) projection involves costly many-point numerical integration~\cite{van2013level}.   Another alternative from the level-set literature is to use sigmoid projection with a level-set function $\phi$ that is constrained to be a signed-distance function ($\Vert \nabla \phi \Vert = 1$), which bounds the width of the intermediate-density regions around interfaces~\cite{van2013level,Wein2020}.  Unfortunately, ensuring that a general $\phi$ remains a signed-distance function requires either re-initialization schemes~\cite{van2013level} (which complicate gradient-based optimization) or penalty terms (requiring careful tuning).  One can also construct a signed-distance $\phi$ from an explicit shape parameterization (such as cylinder radii or other parameterized curves)~\cite{van2013level}, an approach sometimes called ``feature mapping''~\cite{Wein2020}, but this greatly reduces the degrees of freedom compared to freeform~{\TopOpt}.}


\section{Review of density-based topology optimization \label{sec:background}}
Conceptually, density-based topology optimization ({\TopOpt}) maps some array of density values to a real-space grid that is discretized in some fashion (e.g.~finite differences, finite elements). Each pixel within this array is then continuously optimized by interpolating between two or more constituent materials. To ensure the final design is binary (and thus, physically realizable), the threshold condition ($\beta$) of a nonlinear projection function is gradually increased. The canonical optimization problem explored throughout this work corresponds to a frequency-domain, differentiable maximization problem over a single figure of merit (FOM) $f(\vect{E})$:
\begin{equation}
    \begin{matrix}
    & \max_{\vectrho} \mathrm{FOM}(\DOF, \vect{E}_1,\ldots,\vect{E}_n) & \\\ 
    s.t. & \maxwell{n} & n\in\left\{1, 2, \dots, N\right\}\\ &0\le\vectrho\le1&\ \\\
    \end{matrix} \, ,
\end{equation}
where $\DOF$ is the spatially varying material ``density'' (parameterized by the design degrees of freedom or ``DOFs'' $\vectrho$), $\vect{E}_n$ is the time-harmonic electric-field solution at frequency $\omega_n$, $\vect{\epsilon}_r$ is the relative permittivity as a function of the density $\rho$ at each point in space, and $\vect{J}_n$ is the time-harmonic current density (the source term). While we solve problems constrained by Maxwell's equations throughout this paper, we emphasize that our new smoothed-projection approach is agnostic to the underlying physics and applicable to any {\TopOpt} problem.  Adjoint methods (also known as ``backpropagation'' or ``reverse-mode'' differentiation) allow one to compute the gradient of any FOM with respect to \emph{all} DOFs using just two Maxwell solves (the ``forward'' and ``adjoint'' problems)~\cite{Molesky2018,Jensen2010}.

As discussed earlier, to map the ``latent'' design variables ($\DOF$) to the permittivity profile ($\varepsilon$), one first filters the design variables such that
\begin{equation}
    {\DOFfiltered}=\kfilt(\vect{x})*\rho,
\end{equation}
where $\DOFfiltered$ is the filtered design field, $\kfilt$ is the filter kernel, and $*$ is a problem-dependent multidimensional convolution~\cite{projection_overview,svanberg_harmonic}.
The filter radius $\Rfilt$ (or bandwidth $\sim 1/\Rfilt$) is typically determined by the minimum desired lengthscale in the design, and is much larger than the underlying grid/mesh spacing~$\Delta x$---this regularizes the optimization problem to ensure convergence with resolution and is also related to enforcement of fabrication constraints~\cite{zhou_geometric,Hammond21DRC}. Alternatively, one can omit an explicit filtering step by representing the degrees of freedom directly in terms of $\DOFfiltered$, represented by some smooth/band-limited basis/parameterization~\cite{White2018,Sanders2021,Chandrasekhar2022}.

Next, the filtered design variables are projected onto a nearly binary density field, $\DOFprojected$, using \newtext{a} sigmoid \newtext{(smoothed Heaviside)} projection function, $P_{\beta,\eta}(\vect{\DOFfiltered})$, \newtext{such as the following}:
\begin{equation}
    \DOFprojected = P_{\beta,\eta}(\DOFfiltered) = \frac{\rm tanh \left(\beta\eta\right)+\rm tanh \left(\beta\left(\vect{{\DOFfiltered}}-\eta\right)\right)}{\rm tanh\left(\beta\eta\right)+\rm tanh\left(\beta\left(1-\eta\right)\right)},
    \label{eq:tanh}
\end{equation}
where $\beta$ controls the \emph{strength} of the binarization ($\beta\to\infty$ corresponds to a Heaviside step function of~$\DOFfiltered - \eta$, while $\beta \to 0^+$ gives the identity mapping $\DOFprojected \to \DOFfiltered$) and $\eta$ describes the \emph{location} of the threshold point (the level set $\DOFfiltered = \eta$). Finally, these projected densities are linearly interpolated between two (possibly anisotropic) material tensors $\vect{\varepsilon_0}$ and $\vect{\varepsilon_1}$ at each point $\vect{x}$, most commonly via
\begin{equation}
\label{eq:lin_interp}
   \varepsilon(\DOFprojected(\vect{x})) = \DOFprojected(\vect{x})\vect{\varepsilon_0} + (1-\DOFprojected(\vect{x}))\vect{\varepsilon_1}.
\end{equation}
Such linear interpolation between two materials works well if the materials are both non-metallic (positive $\varepsilon$). Different interpolation techniques are employed when optimizing with metals to avoid nonphysical zero crossings in the interpolated $\varepsilon$~\cite{rasmus_interp}, and more generally any differentiable pointwise mapping from $\DOFprojected$ to the material properties could be employed.

\newtext{(As an alternative to the filter/project process, some authors have proposed using a single nonlinear filter, e.g.~morphological transformations, to accomplish both tasks at once~\cite{svanberg_harmonic,Hassan2017,Schevenels2016morph}.  However, since the exact binary-output transformations are not differentiable, one must still introduce a regularization parameter that is gradually reduced, analogous to~$\beta$.)}

Optimization typically proceeds as follows: (i) the design is initialized with some guess (e.g.~a uniform ``gray'' region $\DOF = 0.5$) and a relatively low value for $\beta$; (ii) an optimization algorithm iteratively updates this design according to the value of the FOM and its gradient at each step; (iii) once optimization is sufficiently converged~\cite{Dunning2025}, $\beta$ is increased and the optimization algorithm is restarted using the best performing density from the previous ``epoch''. This process is repeated until the final design is acceptably binary and the performance meets the desired specifications. This ``$\beta$-continuation'' procedure is highly problem dependent, requiring significant effort and experience-gained heuristics to avoid slow convergence, \newtext{although there have been recent efforts to improve automation~\cite{Dunning2025}}. If one increases $\beta$ too rapidly, then the optimization problem quickly becomes stiffer and stiffer, \newtext{necessitating small step sizes~\cite{Guest2011,Dunning2025}} and slowing down the overall convergence. Conversely, if one evolves $\beta$ too slowly, then finding a high-performing binary design is \emph{also} computationally expensive because of the large number of optimization epochs.

Importantly, when differentiating through the physics of the problem, one first computes the gradient with respect to $\DOFprojected$ (which directly determines the physical parameters), rather than $\vectrho$ (the design variables). As such, one must use the chain rule (``backpropagation'') to compute the gradient with respect to $\vectrho$, which is needed by the optimization algorithm. Consequently, the step-function projection is a problem: \newtext{as depicted in \figref{fig:convergence}(b),} the resulting gradient approaches zero almost everywhere as $\beta\to\infty$ (and becomes non-differentiable at the step itself). In other words, standard density-based {\TopOpt} methodologies cannot produce useful gradients for \emph{purely} binary designs. The amount of binarization is directly related to the value of $\beta$, but the designer must avoid making this value too large (regardless of the underlying schedule); at the final step, if $\beta$ is sufficiently large, the structure can hopefully be ``rounded'' to the closest binary structure without greatly degrading performance. In contrast, level-set methods avoid this issue by establishing sophisticated methods to differentiate with respect to an \emph{interface}, which becomes especially cumbersome when the topology changes (e.g.~when interfaces cross). By introducing subpixel smoothing into the density-based formulation, we will remove the non-differentiable discontinuity induced by the projection function and allow the designer to arbitrarily specify~$\beta$.

\section{Efficient subpixel-smoothed projection}
\label{sec:smoothed_projection}

\begin{figure}
    \centering
    \includegraphics{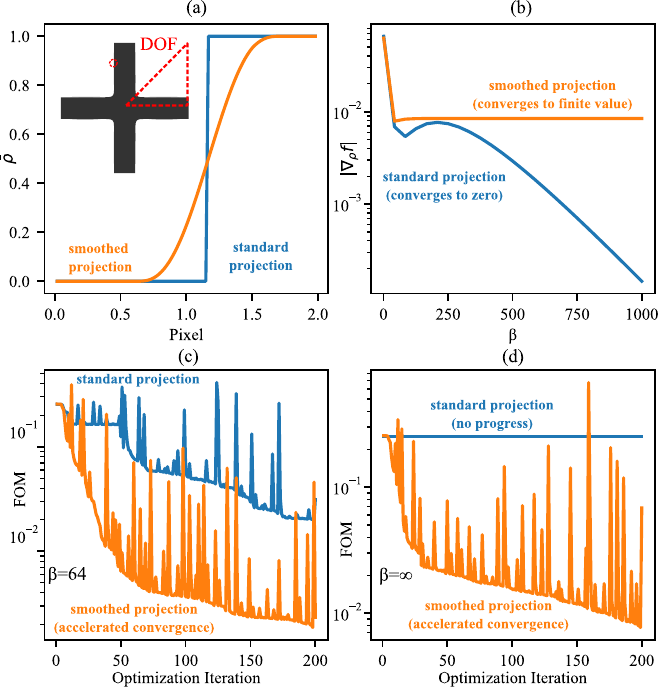}
    \caption{Comparison of the new subpixel-smoothed projection $\DOFsmooth$ (orange) to standard projection $\DOFprojected$ (blue), for optimizing transmission through a waveguide crossing (inset, \secref{sec:crossing}). (a)~Projected densities vs.~position in the vicinity of an interface (small red circle, inset): for $\beta =\infty$, standard projection $\DOFprojected$ is a discontinuous step function, whereas the new $\DOFsmooth$ rises continuously over a 1-pixel ($\sim \Delta x$) distance.  (b)~Gradient norm $\Vert \nabla_\DOF \text{FOM}\Vert$ of the optimization figure of merit (FOM) as a function of projection steepness~$\beta$: for $\DOFprojected$, the gradient is (almost everywhere) zero as $\beta \to \infty$, preventing optimization from making progress; for the new~$\DOFsmooth$, the gradient converges to a nonzero value.  (c)~Optimization progress (FOM vs.~iteration) for the waveguide crossing (\secref{sec:crossing}) with \newtext{$\beta = 64$}: the standard projection converges much more slowly than smoothed projection, because the former is ``stiff'' (ill-conditioned second derivatives) for large $\beta$.  (d)~Optimization progress for $\beta = \infty$: standard projection makes no progress (due to zero gradients), whereas smoothed projection converges at almost the same rate as~(c).
    }
    \label{fig:convergence}
\end{figure}

As $\beta \to \infty$, $\DOFprojected$ becomes a discontinuous function of $\DOFfiltered$, and even for large finite $\beta$ one encounters ill-conditioning problems due to the large second derivatives.  These issues are closely related to the fact that $\DOFprojected$ is a discontinuous function of \emph{space} $\vect{x}$, \newtext{as depicted in \figref{fig:convergence}(a)}: as $\DOFfiltered$ changes, the location of the level set $\DOFfiltered = \eta$ changes smoothly (except at changes in topology), but the discontinuity of $\DOFprojected$ in space means that this smooth interface motion translates to a discontinuous dependence $\DOFprojected(\DOFfiltered(\vect{x}))$ at any given $\vect{x}$.   However, if we could smooth the discontinuities in space---but only near the interfaces so that the projection remains mostly binary---then $\DOFprojected$ would become a differentiable function of $\DOFfiltered(\vect{x})$ due to the smooth interface motion (except at topology changes, discussed in \secref{sec:topology_changes}), even for $\beta = \infty$. This strategy corresponds \emph{conceptually} to a \emph{post}-projection smoothing, but in practice must be implemented in \emph{conjunction} with projection as explained below.

That is, we could conceptually remove the $\DOFprojected$ discontinuity (in space, and hence in $\DOFfiltered$) by convolving $\DOFprojected$ with a smoothing kernel, \emph{if} this convolution could somehow be evaluated at \emph{infinite} spatial resolution, as depicted by the dashed  ``not practical'' arrow in \figref{fig:overview}.  That is, imagine that we could evaluate the exact convolution ($*$) over the computational domain $\Omega$:
\begin{equation}
\DOFsmooth(\vect{x}) = k * \DOFprojected = \int_\Omega \ksmooth(\vect{x}-\vect{x}') \DOFprojected(\vect{x}') \d\Omega' \, ,
\label{eq:DOFsmooth_convolution}
\end{equation}
where $\ksmooth(\vect{x}) \ge 0$ is a localized (radius-$\Rsmooth$) smoothing kernel, with $\int k \d\Omega = 1$ and $\ksmooth = 0$ for $\|\vect{x}\| > \Rsmooth$, and where the support diameter $2\Rsmooth \gtrsim \Delta x$ is proportional to the spacing/resolution $\Delta x$ of our computational grid or mesh (below, we chose $\Rsmooth = 0.55 \Delta x$ to make it slightly larger than a voxel).    (Because $\DOFsmooth$ will be sampled onto a computational grid or mesh, or at quadrature points in a finite-element method~\cite{jin2015finite}, we want $2\Rsmooth$ to be slightly greater than the maximum sample spacing $\Delta x$ to avoid ``frozen'' gradients where the derivative is zero at all sampled~$\vect{x}$.)  If we could do this, then $\DOFsmooth$ would have two desirable properties: (i) it would remain~$0$ or~$1$ as~$\beta \to \infty$ except in a radius-$\Rsmooth$ (``subpixel'') neighborhood of interfaces (i.e.~binary \emph{almost everywhere} as the computational resolution increases: $\Delta x \to 0$); and (ii) if $\ksmooth$ is sufficiently smooth, then $\DOFsmooth$ will be a differentiable function of $\vect{x}$ and also of the underlying $\DOF$ even for $\beta \to \infty$ (as long as the position of the $\DOFfiltered = \eta$ level set changes smoothly with $\DOF$).  
The challenge is to make the computation of such a $\DOFsmooth$ \emph{practical and efficient}.

\begin{figure}
    \centering
    \includegraphics{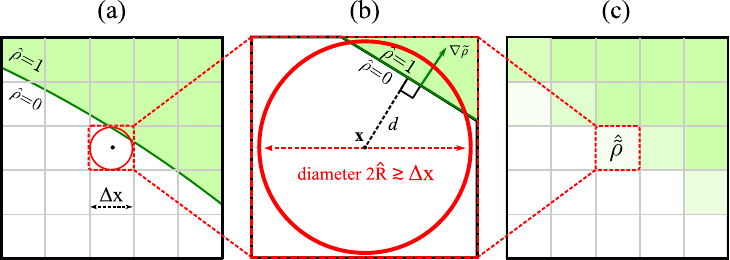}
    \caption{Schematic of proposed subpixel smoothing routine, shown here in the $\beta \to \infty$ limit (step-function projection $\DOFprojected$). A simulation algorithm discretizes the geometry on some grid or mesh with spacing $\sim \Delta x$ (a). At a particular grid point $\vect{x}$, a smoothing sphere is chosen with diameter $2\Rsmooth \gtrsim \Delta x$ (b).  In a small region, the interface is approximately planar ($\perp \nabla \DOFfiltered$), at a distance~$d$ from~$\vect{x}$.  Conceptually, a subpixel-smoothed $\DOFfiltered$ is constructed by convolving $\DOFprojected$ with some smoothing kernel of radius $\Rsmooth$, resulting in an almost-every binary discretized structure~(c) ($\DOFsmooth \in \{0,1\}$ for $d \ge \Rsmooth$).  To make this practical and differentiable, we compute $\DOFsmooth$ from $\DOFfiltered$ instead of from $\DOFprojected$.}
    \label{fig:schematic}
\end{figure}

The key fact is that the underlying filtered field $\DOFfiltered$ is, by construction, smooth and slowly varying (on the scale of the filter radius $\Rfilt \gg \Rsmooth \sim \Delta x / 2$), which allows us to smoothly interpolate it to any point $\vect{x}$ in space---thus implicitly defining $\DOFprojected(\vect{x}) = P_{\beta,\eta}(\DOFfiltered(\vect{x}))$ at infinite resolution---and furthermore it implies that $\DOFfiltered$ can normally be \emph{linearized} in a small neighborhood $\| \vect{x}' - \vect{x} \| \le \Rsmooth \ll \Rfilt$:
\begin{equation}
\DOFfiltered(\vect{x}') \approx \DOFfiltered(\vect{x}) + \left. \nabla \DOFfiltered \right|_{\vect{x}} \cdot (\vect{x}'-\vect{x}) \, .
\label{eq:DOF_linearized}
\end{equation}
This allows us to vastly simplify the smoothing computation of~$\DOFsmooth$ as explained below, and will ultimately allow us to completely replace the integral~\eqref{eq:DOFsmooth_convolution} with a simple local function of~$\DOFfiltered$ and~$\Vert \nabla \DOFfiltered \Vert$.   Moreover, in the linearized approximation, the signed distance~$d$ from any point $\vect{x}$ to the (approximately planar) $\DOFfiltered = \eta$ level set (i.e., the $\DOFprojected$ discontinuity at $\beta=\infty$) is simply:
\begin{equation}
d = \frac{\eta - \DOFfiltered(\vect{x})}{\left\Vert \left. \nabla \DOFfiltered \right|_{\vect{x}} \right\Vert} \, ,
\label{eq:d}
\end{equation}
as depicted in \figref{fig:schematic}.  Note that we define $d > 0$ or $d < 0$ for $\DOFfiltered$ below or above $\eta$, corresponding to
 $\DOFprojected(\vect{x}) = 0$ or~$1$ for $\beta\to\infty$, respectively.  (For cases where optimization causes two interfaces to meet, changing the topology of the level set, the effect of this linearization is discussed in \secref{sec:topology_changes}.)

The next step is to choose the smoothing kernel to be rotationally symmetric $\ksmooth(\vect{x}) = \ksmooth(\| \vect{x} \|)$, and to rewrite the convolution (\ref{eq:DOFsmooth_convolution}) in spherical coordinates.  (Henceforth, we will always view the convolution as being performed in~3d: even if the underlying problem is~1d or~2d, we will treat it as conceptually ``extruded'' into 3d, e.g.~a 2d $\DOF(x_1,x_2)$ is treated as a 3d $x_3$-invariant function.) In particular, we center the spherical coordinates at $\vect{x}$, where the ``$z$'' axis ($\theta = 0$, $z \ne x_3$) is oriented in the $\nabla \DOFfiltered$ direction, and approximate the integral~\eqref{eq:DOFsmooth_convolution} in terms of the linearized $\DOFfiltered$ from \eqnref{eq:DOF_linearized}:
\begin{equation}
\DOFsmooth(\vect{x}) \approx 2\pi \int_0^{\pi} \int_0^{\infty} \ksmooth(r) P_{\beta,\eta}(\DOFfiltered(\vect{x}) + \| \nabla \DOFfiltered \| r  \cos\theta) \, r^2 \sin(\theta) \d r \d \theta \, .
\label{eq:DOFsmooth_linearized_conv}
\end{equation}
In principle, \eqnref{eq:DOFsmooth_linearized_conv} is already computationally tractable: at each point $\vect{x}$ in our grid/mesh where we want to compute $\DOFsmooth$, we evaluate $\DOFfiltered$ and $\nabla \DOFfiltered$, and then apply a numerical-quadrature routine to evaluate the integral.   This is especially cheap for $\beta \to \infty$ because the integral is simply $1$ or $0$ except for points $\vect{x}$ where $|d| < \Rsmooth$, so quadrature need only be employed at a small set of points near the interface.   However, we can make further efficiency gains by the choice of kernel~$\ksmooth$ and other approximations described below.

First, consider the limit $\beta \to \infty$, for which $P_{\infty,\eta}$ is a Heaviside step function.  Suppose that the point $\vect{x}$ is within $|d| < \Rsmooth$ of the $\DOFprojected$ interface, and is on the $\DOFprojected = 0$ side ($d \ge 0$) as in \figref{fig:schematic}.  In this case, the integral~(\ref{eq:DOFsmooth_linearized_conv}) simplifies even further to a function $F(d)$ that depends on $d$ alone:
\begin{equation}
F(d) = 2\pi \int_0^{\pi/2}  \underbrace{\int_{d\sec\theta}^{\infty} \ksmooth(r) \, r^2 \sin(\theta) \d r}_{\DOFprojected=1 \mbox{ region}} \d \theta \, .
\label{eq:F}
\end{equation}
For any given kernel $\ksmooth(r)$, this function $F(d)$ could be precomputed.   If the kernel $\ksmooth$ is simply a ``top-hat'' filter $\ksmooth(r) = \nicefrac{3}{4\pi \Rsmooth^3}$ for  $r \le \Rsmooth$ and $\ksmooth(r)=0$ for $r > \Rsmooth$, then $F(d)$ is a ``fill factor'' of the 
$\DOFprojected = 1$ region inside the sphere, and can be computed analytically.  For $d < 0$, we let $F(-d) = 1 - F(d)$: the fill factor is inverted (equivalent to inverting $\DOFprojected \to 1 - \DOFprojected$).  \newtext{(On a uniform mesh, another possibility would be a $\hat{k}$ given by the indicator function of the mesh elements, e.g.~$1/\Delta x^3$ inside a cubic voxel for a Cartesian mesh, which would be closely analogous to fill-fraction level-set methods~\cite{deRuiter2004,van2012explicit,van2013level}. As noted in \secref{sec:intro}, this would make $\DOFsmooth$ much harder to calculate due to the broken symmetry, and has potentially discontinuous derivatives.)}   However, rather than specifying $\ksmooth(r)$ and computing $F(d)$, it is attractive to instead choose~$F(d)$ and infer~$\ksmooth(r)$, because $F(d)$ directly determines the \emph{smoothness} of $\DOFsmooth$ as a function of $\DOF$ (recalling that $d$ is a smooth function of $\DOFfiltered$, which in turn is a smooth function of $\DOF$).  The algebraic details are given in \appenref{A}, where we construct an inexpensive, twice-differentiable, monotonic, piecewise-polynomial choice of $F$, denoted as~$F_2(d)$:
\begin{equation}
F_2(d) = \begin{cases}
\frac{1}{2} - \frac{15}{16} \frac{d}{\Rsmooth} + \frac{5}{8} \left(\frac{d}{\Rsmooth}\right)^3 - \frac{3}{16}\left(\frac{d}{\Rsmooth}\right)^5 & |d| \le \Rsmooth \\
0 & d > \Rsmooth \\
1 & d < -\Rsmooth
\end{cases} \, ,
\label{eq:Fpoly}
\end{equation}
which is continuous with a continuous derivative and second derivative, satisfies $F_2(-d)=1-F_2(d)$, and turns out to correspond to a parabolic kernel $\ksmooth_2(r)$ proportional to $1 - (r/\Rsmooth)^2$ for $r \le \Rsmooth$.  We also show how to construct an infinitely differentiable $F$ if desired: see \eqnref{eq:Fsmooth}.

Thus, for $\beta \to \infty$, this $\DOFsmooth$ function is twice differentiable (in $d$ and hence $\DOF$) by \emph{construction} (except at changes in topology as discussed in \secref{sec:topology_changes}), is almost-everywhere binary (in the limit of $\Delta x \to 0$), and is notably cheap to compute at any desired point $\vect{x}$ in space:  \begin{enumerate}
    \item Given the smooth $\DOFfiltered$ function, interpolate it and its gradient to $\vect{x}$ (e.g.~via bilinear interpolation in a gridded finite-difference method),

    \item Compute $d$ by \eqnref{eq:d}, 

    \item Compute $\DOFsmooth(\vect{x}) = F(d)$ for the chosen $F$, e.g.~\eqnref{eq:Fpoly} or \eqref{eq:Fsmooth}.  This is generalized to \eqnref{eq:DOFsmooth_beta} for finite~$\beta$ below.
\end{enumerate}
Each of these steps is differentiable (except as noted in \secref{sec:topology_changes}), so backpropagating derivatives to $\DOF$ for adjoint differentiation is straightforward.  Much like the computation of $\DOFprojected$ in standard {\TopOpt}, $\DOFsmooth$ is a purely local function of $\DOFfiltered$ at each $\vect{x}$, with the only difference being that it now depends on both $\DOFfiltered$ and $\Vert \nabla\DOFfiltered \Vert$.  (In a discretized grid/mesh, such local quantities are determined by interpolating adjacent mesh points.)  Moreover, even though $\DOFsmooth \to \DOFprojected$ as $\Delta x \to 0$, it remains a well-conditioned function of the~$\DOFfiltered$ values (it has bounded second derivatives) on any discretized mesh, as discussed in \appenref{B} \newtext{and \secref{sec:convergence}}.

\begin{figure}
    \centering
    \includegraphics[width=0.7\columnwidth]{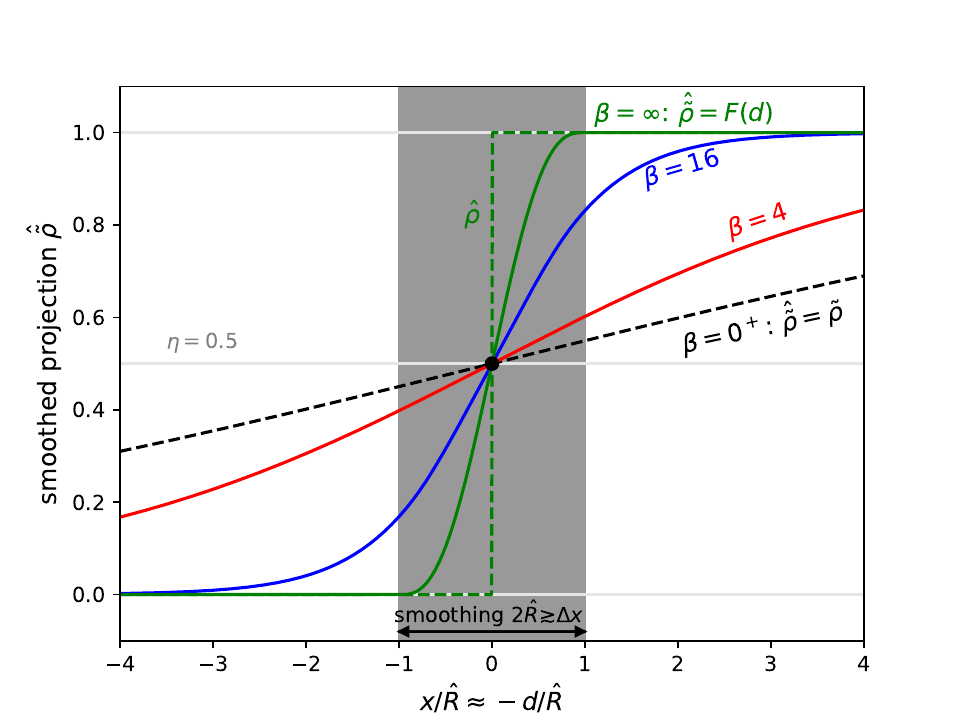}
    \caption{Subpixel-smoothed projections $\DOFsmooth(x)$, \eqnref{eq:DOFsmooth_beta}, applied to an example $\DOFfiltered(x) = \frac{1}{2} \tanh(x/\Rfilt) + \frac{1}{2}$ (dashed black) for $\Rfilt = 10\Rsmooth$, with a threshold $\eta = 0.5$, for various steepness parameters $\beta$.  ($\DOFfiltered$ is nearly linear for $-4 < x/\Rsmooth < 4$, so \eqnref{eq:d} gives $d \approx -x$.)  The shaded region indicates the smoothing radius~$\Rsmooth$ around the level set~$\DOFfiltered = \eta$.  For $\beta=0^+$ (dashed black), our formula is equivalent to no projection ($\DOFsmooth = \DOFfiltered$), while for $\beta = \infty$ (solid green) we obtain our chosen ``fill-fraction'' curve $F(d)$; for reference, the step-function projection $\DOFprojected$ for $\beta = \infty$ is also shown (dashed green).    Here, we use $F = F_2$ from \eqnref{eq:Fpoly}, so the $\DOFsmooth$ curves are all twice differentiable by construction, despite the fact that the piecewise definitions change at $d/\Rsmooth=\pm 1$.}
    \label{fig:smoothed-projection-approx}
\end{figure}

For finite~$\beta$, one could fall back to numerical integration of \eqnref{eq:DOFsmooth_linearized_conv}.  However, we can greatly speed up the finite-$\beta$ computation by making two approximations, which still result in a differentiable $\DOFsmooth$ that reproduces the convolution \eqref{eq:DOFsmooth_linearized_conv} above as $\beta \to \infty$.   First, for points $|d| \ge \Rsmooth$ far from the level set, we simply omit the smoothing and choose $\DOFsmooth(\vect{x}) = \DOFprojected(\vect{x})$ as in conventional {\TopOpt}: far from an interface, the projection $\DOFprojected$ is nearly constant within the smoothing radius $\Rsmooth$,  so there is not much point in smoothing it further.  Second, for points $|d| < \Rsmooth$ close to the level set, we employ the $\beta \to \infty$ fill-fraction function $F(d)$ as above, with the modification that we average  values of $\DOFprojected$ evaluated on the two sides of the level set, denoted by $\DOFprojected_{-}$ and $\DOFprojected_{+}$ and defined below, instead of averaging~$0$ and~$1$.  Rather than an approximation for \eqnref{eq:DOFsmooth_linearized_conv}, this can be viewed as simply a new \emph{definition} of $\DOFsmooth$, namely:
\begin{equation}
\DOFsmooth(\vect{x}) =
\begin{cases} 
  (1 - F(d)) \DOFprojected_{-} + F(d)\DOFprojected_{+}, & \text{if } |d| < \Rsmooth \\
  \DOFprojected(\vect{x}), & \text{otherwise},
\end{cases}
\label{eq:DOFsmooth_beta}
\end{equation}
where $\DOFprojected_{\pm}$ are obtained by projecting the linearized approximation of $\DOFfiltered$ at two points along the~$\nabla\DOFfiltered$ direction (so that they still depend only on $\DOFfiltered$ and $\Vert\nabla\DOFfiltered\Vert$ at $\vect{x}$) within~$\Rsmooth$ of~$\vect{x}$:
\begin{equation}
\DOFprojected_{\pm}(\vect{x}) =
P_{\beta,\eta} \left( \, \DOFfiltered(\vect{x}) \pm \Rsmooth \, \Vert\nabla\DOFfiltered\Vert \, F(\mp d)\, \right) \, ,
\label{eq:rhohat_pm}
\end{equation}
corresponding to two points on either side of the level set separated by a distance $\Rsmooth F(-d) + \Rsmooth F(d) = \Rsmooth$.
 This $\DOFsmooth$, demonstrated in \figref{fig:smoothed-projection-approx}, is constructed to make \eqnref{eq:DOFsmooth_beta} monotonic and twice differentiable in $d$, and also so that \eqnref{eq:DOFsmooth_beta} reduces to $\DOFsmooth \to F(d)$ (as above) in the limit $\beta \to \infty$ (in which case $\DOFprojected_- \to 0$ and $\DOFprojected_+ \to 1$) as well as to $\DOFsmooth \to \DOFfiltered$ as $\beta \to 0^+$ (no projection), as can be shown via straightforward algebra.   In particular, note that \eqnref{eq:rhohat_pm} re-uses our function $F(d)$ purely for convenience, exploiting the fact that $F$ goes smoothly from~$1$ at~$d=-\Rsmooth$ to~$0$ at $d=\Rsmooth$ to make \eqnref{eq:DOFsmooth_beta}  transition smoothly to the~$\DOFprojected(\vect{x})$ case as~$d \to \pm\Rsmooth$.

Our new subpixel-smoothed projection $\DOFsmooth$ \eqref{eq:DOFsmooth_beta} is computationally inexpensive while offering two fundamental benefits over traditional projection \eqref{eq:tanh}: (i) the ability to optimize even as $\beta\to\infty$; and (ii) improved convergence for large finite values of $\beta$ (because it has a bounded second derivatives regardless of~$\beta$, from~\appenref{B}, leading to better-conditioned optimization problems~\cite{bertsekas2016optimization}) and more rapid convergence for large~$\beta$.  
In \secref{sec:experiments}, we quantify these advantages with numerical experiments, illustrations of which are shown in \newtext{\figref{fig:diverging_optimization} and \figref{fig:convergence} as discussed in \secref{sec:convergence}.  The almost-everywhere binary result of SSP for $\beta=\infty$ is illustrated in the inset of \figref{fig:diverging_optimization}.}

\subsection{Differentiating through changes in topology?}
\label{sec:topology_changes}

As the density $\DOF$ is optimized, one occasionally encounters situations where the topology (the connectivity and number of holes/islands) of $\DOFprojected$  changes (for $\beta = \infty$), for example when the diameter of some $\DOFprojected=0$ region goes to zero.
In such cases, where \emph{two} interfaces approach within a separation~$\sim \Rsmooth$, the linearized $\DOFfiltered$ of \eqnref{eq:DOF_linearized} is no longer accurate at points adjacent to both interfaces.  For example, at a point midway between two interfaces, $\DOFfiltered$ will reach a local minimum (or maximum) where $\nabla \DOFfiltered$ vanishes, in which case the second derivative of $\DOFfiltered$ cannot be neglected, nor can the interface be treated as a single plane.   (Correspondingly, $d$ passes through $\pm \infty$.) What happens to our $\DOFsmooth$ formulas in such cases?

If we employ the linearization-based construction of \eqnref{eq:DOFsmooth_beta}, the behavior near changes in topology is straightforward and relatively benign: it overestimates $d$ because of the small denominator in \eqnref{eq:d}, so \eqnref{eq:DOFsmooth_beta} reduces to the traditional projection $\DOFprojected$.   Thus, SSP's ability to optimize through changes in topology is \emph{no worse} than in traditional density-based {\TopOpt}: it is differentiable for finite~$\beta$.   For $\beta \to \infty$, with \eqnref{eq:DOFsmooth_beta} we can still perform ``shape optimization'' (fixed topology) as demonstrated in~\secref{sec:experiments}, but a discontinuity occurs for changes in topology.   Thus, in order to perform full topology optimization, we still follow the traditional strategy~\cite{Wang2011,projection_overview,meep_adjoint} of optimizing over a sequence of increasing~$\beta$ values, except that now we can increase $\beta$ much more quickly (e.g.~$\beta = 16, 32, \infty$) as long as the finite-$\beta$ calculations first obtain the desired topology so that the $\beta=\infty$ optimization need only adjust the shape. \newtext{(Doing a few steps of finite-$\beta$ optimization also has the benefit that new holes/islands can nucleate anywhere, whereas for $\beta=\infty$ the structure can only evolve from the boundaries.)}

Moreover, our formulation offers a future pathway to  performing {\TopOpt} directly for $\beta=\infty$ as well, without requiring cumbersome topological derivatives.  By monitoring the second derivative of $\DOFfiltered$ (e.g.~using bicubic interpolation), one can determine when the linearization~\eqref{eq:DOF_linearized} fails, and for those points $\vect{x}$ one could simply switch to a more complicated smoothing procedure.  Since such points, where the topology is changing, will occur over only a tiny subset of the design region, the expense of even a brute-force convolution/quadrature at those points should be negligible.   We plan to explore this possibility in subsequent work as discussed in \secref{sec:conclusion}, but for the present even the simplified SSP formulation of \eqnref{eq:DOFsmooth_beta} appears to be greatly superior to traditional projection.

\section{Numerical experiments}
\label{sec:experiments}

Here, we present numerical experiments using three different Maxwell discretization techniques to illustrate the flexibility of our new SSP method. First, we design a 2d silicon-photonic crossing using a hybrid time-/frequency-domain adjoint solver~\cite{meep_adjoint} compatible with Meep, a free and open-source finite-difference time-domain~(FDTD) code (FDTD)~\cite{meep}.  The SSP formula~\eqref{eq:DOFsmooth_beta} is implemented as a vectorized function in Python, automatically differentiable via the autograd software~\cite{maclaurin2015autograd}, given in Appendix~C. We demonstrate that SSP is directly compatible with shape optimization (where an initial design topology is already identified and optimized at $\beta=\infty$) \emph{and} with freeform {\TopOpt} (at finite~$\beta$ then $\beta=\infty$ as discussed in \secref{sec:topology_changes}) on an example 2d waveguide-crossing~\cite{wu2020state} design problem.  We also show greatly accelerated convergence compared to traditional density projection~\eqref{eq:tanh} for a large finite $\beta=32$.

Next, we illustrated the versatility of SSP by applying it to other problems taken from our photonics-optimization test suite~\cite{testbed}, as well as to other computational-electromagnetics methods.
We design a 3d diffractive metasurface using FMMAX~\cite{fmmax}, a differentiable and GPU-accelerated Fourier Modal Method (FMM) code written in Python. We then design a 2d metallic nanoparticle to enhance near-field concentration using a finite-element electromagnetics solver implemented in the Julia software package Gridap~\cite{Badia2020}, demonstrating that the same method is applicable to discretizations employing unstructured meshes.

\subsection{FDTD: Integrated Waveguide Crossing}\label{sec:crossing}

The goal of the silicon photonics waveguide crossing is to maximize transmission from one side to the other while minimizing crosstalk into the orthogonal waveguides. Here, the design region is 3~$\mu$m $\times$~3~$\mu$m, with $\DOF$ discretized at a resolution of 60~pixels/$\mu$m, twice the FDTD simulation resolution (30~pixels/$\mu$m) \newtext{to correspond to the underlying Yee grid}. An eigenmode source injects the fundamental $E_z$-polarized (out-of-plane) mode from the left, and a monitor records the fraction of power leaving the waveguide on the right in this same mode. The amplitude $\alpha_m^{\pm}$ of an outgoing mode (power $\sim |\alpha_m^{\pm}|^2$) is given by overlap integral~\cite{snyder_love}.
\begin{equation}
    \alpha_m^{\pm} = \int_A \left[\vect{E}^*(r)\times\vect{H}_m^{\pm}(r)+\vect{E}_m^{\pm}(r)\times\vect{H}^*(r)\right]\cdot\vect{\hat{n}} \d A
    \label{eq:mode_overlap}
\end{equation}
for the $m^{\mathrm{th}}$ mode for the forward ($+$) and backward ($-$) directions, $\vect{E}(r)$ and $\vect{H}(r)$ are the Fourier-transformed total (simulated) fields at a particular frequency, $\vect{E}_m^{\pm}(r)$ and $\vect{H}_m^{\pm}(r)$ are the eigenmode profiles (normalized to unit power) at the same frequency for the forward- ($+$) and backward- ($-$) propagating modes, and $A$ is a waveguide cross-section.

To simplify the optimization problem, we explicitly enforce horizontal, vertical, and $C^4$-rotational symmetries using differentiable transformations on the design degrees of freedom before filtering and SSP-projecting them, so that the crossing works well in all four directions.
(Separate optimization constraints intended to minimize power in the orthogonal waveguides are not needed because optimization attains nearly $100$\% transmission.) The resulting optimization problem is described by
\begin{equation}
    \begin{matrix}
     & \max_{\vectrho} \frac{|\alpha_0^{+}|^2}{P_0} & \\\ 
    \mathrm{s.t.} & \maxwell{}
    \\ &0\le\vectrho\le1&\ \\\ 
    \end{matrix},
\end{equation}
where $|\alpha_0^{+}|^2$ is the power propagating in the fundamental mode through the output waveguide and normalized by the total input power, $P_0$. For these experiments, we optimize at a single wavelength, $\lambda=1.55 \mu$m (in which case it is known that nearly 100\% transmission can be achieved via resonant effects~\cite{JohnsonMa98}).

\begin{figure}
    \centering
    \includegraphics{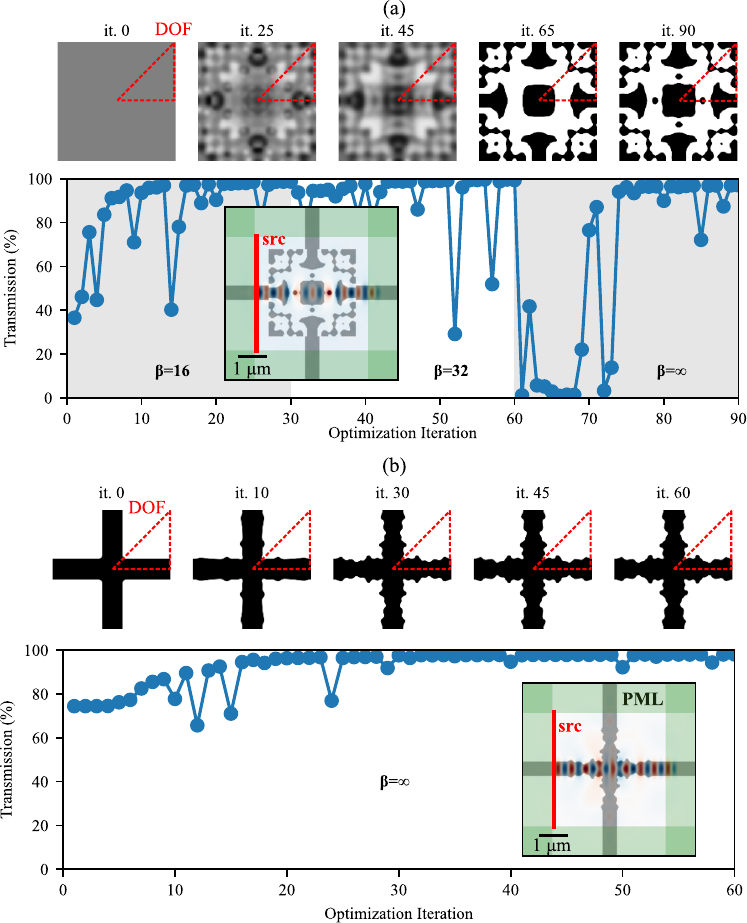}
    \caption{Transmission FOM vs.~iteration for optimizing a silicon ($\varepsilon =12$) waveguide crossing, using subpixel-smoothed projection and FDTD~\cite{meep, meep_adjoint} simulations. (a)~Topology optimization by increasing $\beta$ during optimization ($\beta=16,32, \infty$) starting with $\DOF = 0.5$ (it.~0 inset). (b)~Shape optimization ($\beta=\infty$, fixed topology) starting with a naive crossing (it.~0 inset). 
    Upper insets show sequence of optimized shapes at different iterations (it.), where the DOF are restricted to a triangle (red dashed) to impose four-fold symmetry.  Lower insets show entire computational cell (including PML absorbing boundaries) and simulated electric field $E_z$ (red/blue = postive/negative) for the final structure.}
    \label{fig:waveguide_TO}
\end{figure}

First, we design the crossing using density-based {\TopOpt}. A 2d array of density voxels ($\DOF$) is uniformly initialized to 0.5 (gray) before applying the symmetry, filtering, and SSP transformations as described above. We apply the CCSA optimization algorithm~\cite{Svanberg2002} implemented within the NLopt library~\cite{nlopt} for a sequence $\beta=16, 32, \infty$ of $\beta$ values, for 30 iterations at each $\beta$. \figref{fig:waveguide_TO}(a) shows the geometric and performance evolution of the optimization. As expected, the optimizer produces a strictly binary device that exhibits high transmission from left to right, with a freeform structure entirely different from the initial conditions, exhibiting several disconnected regions. One common phenomenon typical of topology optimization is the sudden drop in performance when the projection parameter changes from $\beta=32$ to $\beta=\infty$: the resulting change in geometry is so drastic that all the performance gains from the previous iterations are apparently lost. To combat these effects, designers often tune a ``schedule'' for gradual increase of~$\beta$, requiring extensive trial and error. In this case, however, gradual increase of~$\beta$ is not needed: the original performance is quickly recovered by shape optimization at $\beta=\infty$, only slightly adjusting the geometry.

Alternatively, we can start with a reasonable initial condition for the design with a chosen topology, and simply optimize that directly via shape optimization at $\beta=\infty$. SSP allows us to use the same machinery as before, but set $\beta=\infty$ from the beginning. In this case the structure was initialized to a ``naive'' waveguide crossing consisting of a binary cross shape (it.~0 in \figref{fig:waveguide_TO}). We ran the optimizer for 60 iterations. \figref{fig:waveguide_TO}(b) illustrates the geometric and performance evolution of this shape optimization.  An enlarged view of the final result, showing the 1-pixel layer of intermediate materials at interfaces, is shown in the inset of \figref{fig:diverging_optimization}.  The resulting structure exhibits performance comparable to that of the topology optimized structure, but the design itself more closely resembles the naive crossing---as expected, the \emph{topology} of the structure remains unchanged (no new holes or islands). As such, techniques like this are best applied when an intelligent starting guess is already known.

\newtext{\subsubsection{Convergence vs. projection steepness~$\beta$}
\label{sec:convergence}

The shape-optimization example of \figref{fig:waveguide_TO}(b) is also useful to illustrate how standard projection leads to slow convergence for large~$\beta$~\cite{Guest2011,Dunning2025} while SSP does not.  \figref{fig:diverging_optimization} shows the number of iterations required to converge to 99\% transmission as a function of~$\beta$: using standard projection, the number of iterations diverges with $\beta$ due to the increasing ill-conditioning; conversely, with SSP the number of iterations remains finite even for $\beta = \infty$.   A detailed view of this faster convergence is shown in \figref{fig:convergence}(c) for $\beta = 64$.  The $\beta=\infty$ convergence history is shown in \figref{fig:convergence}(d): in this case, the piecewise-constant standard projection cannot converge at all in a gradient-based algorithm, whereas SSP still makes rapid progress.

Note that the convergence in \figref{fig:convergence}(c--d) appears irregular because we are also showing the ``inner'' iterations of CCSA, in which the algorithm often aggressively takes too large a step and then backtracks by increasing a penalty~\cite{Svanberg2002}.  This same irregularity also explains why \figref{fig:diverging_optimization} is somewhat nonsmooth/nonmonotonic.   More regular convergence can be obtained by limiting the step size, at the risk of an increased number of iterations.  For example, we also tried the same problem with the more-conservative Adam optimization algorithm~\cite{adam} using its default hyper-parameters, and obtained qualitatively similar results that were more smooth and monotonic, but which required $\sim 10\times$ more iterations (for both SSP and standard projection).}

\subsection{FMM: Metasurface Diffraction Grating}

We now describe the optimization of a 3d metasurface diffraction grating, similar to a number of prior studies~\cite{sell2017large,sell2018ultra,sell2017periodic,jiang2019global,jiang2020simulator}, with the specific parameters taken from our optimization test suite~\cite{testbed} and implemented in the invrs-io gym~\cite{schubert2024invrsgymtoolkitnanophotonicinverse}. The grating consists of a single silicon layer ($n=3.45$) resting on a silica substrate ($n=1.45$), and is designed to deflect normally-incident, linearly $p$-polarized ($E_x, H_y$) light in air ($n=1$) to the +1 diffraction order ($\theta=50^\circ$). The wavelength used is 1050~nm and the thickness of the silicon layer is 325~nm. The grating is periodic in the $x$ and $y$ directions, with a reflection symmetry in the $y-$ direction; the unit cell is 1370~nm $\times$ 525~nm.

We simulated the grating using FMMAX~\cite{fmmax}, a free and open-source implementation of the Fourier modal method (FMM, also known as rigorous coupled wave analysis, RCWA) written in jax~\cite{jax}, which is both GPU-accelerated and differentiable. 

The figure of merit for this problem is the diffraction efficiency: the fraction of the incident power deflected into the desired diffraction order. The resulting optimization problem is described by
\begin{equation}
    \begin{matrix}
     & \max_{\vectrho} \frac{|\alpha_1^{+}|^2}{P_0} & \\\ 
    \mathrm{s.t.} & \maxwell{}
    &0\le\vectrho\le1&\ \\\ 
    \end{matrix},
\end{equation}
where $\alpha_1^{+}$ corresponds to the mode amplitude [similar to \eqnref{eq:mode_overlap}] of the $p$-polarized ($H_y$) $+1$ diffraction order, and $P_0$ is the total injected power from the normally incident planewave.

\begin{figure}
    \centering
    \includegraphics{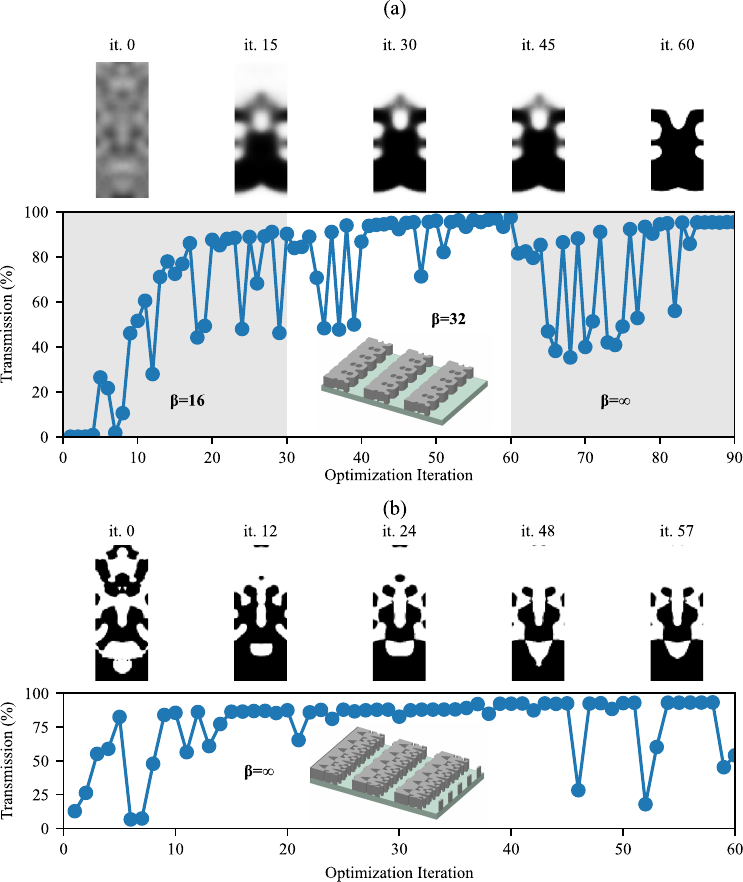}
    \caption{Optimization of a 3d metagrating (insets) maximize transmission of a normal-incident planewave (from below) into the first diffracted order~\cite{testbed}, starting from random structures, using subpixel-smoothed projection and Fourier modal method (FMM)~\cite{kim2012fourier} simulations.  (a) Topology optimization, using $\beta = 16,32,\infty$, halted at 90 iterations with performance $93.2$\%.  (b) Shape optimization ($\beta = \infty$); halted at 60 iterations with final best performance $95.4$\% (from iteration~57), similar to previous work~\cite{sell2017large,sell2018ultra,sell2017periodic,testbed}. }
    \label{fig:metagrating_TO}
\end{figure}

As with the waveguide crossing above, we first optimized the structure using {\TopOpt}. We ran the optimization for 30 iterations per $\beta$, with $\beta= 8,  16, \infty$. We initialized the structure with uniform random $\DOF \in [0,1]$ values at each pixel, and dimensions of 472$\times$180 pixels, projected to mirror symmetry in $y$ (i.e.,~effectively only 90 pixels are free parameters in that direction) \newtext{to halve the computational cost (while still permitting high-performance designs)}. \newtext{(This problem exhibits a large number of local minima with similar performance, and even using a deterministic starting point does not guarantee that the same minimum is found by different algorithms; however, the distribution of minima is reproducible~\cite{testbed}.)} \figref{fig:metagrating_TO}(a) shows the geometric and performance evolution of the diffractive metasurface designed using {\TopOpt}. The final structure yields a diffraction efficiency of 95.4\%, which is within 2\% of the highest performing devices previously reported for this problem~\cite{sell2017large,sell2018ultra,sell2017periodic,testbed}, with a fully binary design whose topology was determined by optimization.  As discussed in \citeasnoun{testbed}, this problem has many local optima, depending on the initial~$\DOF$, but most random starting points converge to a performance $> 90$\%.

Next, we performed shape optimization, using the same initial~$\DOF$, but now projected to a random binary $\DOFsmooth$ via $\beta=\infty$ from the beginning. We ran the optimization for 60 iterations at $\beta=\infty$. \figref{fig:metagrating_TO}(b) shows the geometric and performance evolution. The final structure yields a diffraction efficiency of 93.2\%, which is close to that of the {\TopOpt} result, but represents a different local optimum. 
Shape optimization is more restricted by the topology of the initial guess, although we find it does occasionally change the topology (despite the lack of differentiability at these transitions): some small ``island'' structures disappear, though new islands do not nucleate. 

\subsection{FEM: Near-field Focusing}

Finally, we tested the near-field focusing design problem from \citeasnoun{testbed}, where the objective is to design a metallic structure which enhances the incidence of scattering, such that it \emph{focuses} onto a molecule (located in the center of the structure), inspired by related problems in plasmonic-resonance design~\cite{christiansen2020inverse}. Monochromatic light at 532~nm and polarized in the plane (i.e.,~$H_z$) is incident upon a silver ($n=0.054+3.429i$), 2d device that utilizes combined plasmonic and spatial resonances, with the metal constrained to lie within an annulus (a disc with a hole), to enhance the field at the center. The figure of merit for this problem is to maximize the light intensity at the molecule, and the resulting optimization problem is described by
\begin{equation}
    \begin{matrix} 
     & \max_{\vectrho} \frac{|\boldsymbol{E}(\boldsymbol{x}_0)|^2}{|\boldsymbol{E}^{(0)}(\boldsymbol{x}_0)|^2} & |\boldsymbol{E}|^2 = |\nabla H_z|^2 & \\\ 
    \mathrm{s.t.} & \maxwellH{} 
    &0\le\vectrho\le1&\ \\\ 
    \end{matrix},
\end{equation}
where $H_z$ is the field measured at the molecule, $\boldsymbol{E}^{(0)}$ is the input field (or the field at the molecule in the absence of the metal structure), and $f(\boldsymbol{x})$ is the magnetic current density which for this problem is just an incident plane wave $\boldsymbol{E}^{(0)}$.

To simulate this structure, we used Gridap~\cite{Badia2020}, a free/open-source finite-element method (FEM) implementation in Julia. The corresponding weak form for this problem is defined by
\begin{equation}
    a(u,v,p) = \int_\Omega \left[\nabla(\Lambda v)\cdot\frac{1}{\varepsilon(\DOF)}\Lambda\nabla u-k^2uv\right] \d \Omega,
\end{equation}
where $u$ and $v$ are first-order Lagrangian basis functions, $\ksmooth$ is the normalized wavenumber, and $\Lambda$ is a perfectly matched layer (PML) operator~\cite{taflove_em_book}. A triangular mesh at a resolution of 1~nm was generated to discretize the fields over the design domain, with outer radius $100$~nm.  Although one could discretize $\DOF$ on the same triangular mesh as the fields, we instead chose to discretize $\DOF$ on a Cartesian grid of $532\times 532$~pixels for a $200\times 200$~nm square enclosing the design region; $\DOFfiltered$ is then set to zero (air) outside of the annular design domain.  In center of the design region, the material is constrained to be air within a $10$~nm region: this regularizes the problem by prohibiting sharp tips yielding arbitrarily large fields at the focal spot~\cite{testbed}. Outside the design region, the material is also air, with a mesh resolution of 20~nm. We use PML around the boundaries, and the total simulation region was 800~nm by 600~nm. An incident planewave is generated by a line-current source 200~nm above the top boundary.

From integrating the weak form for each pair of basis functions, one constructs a ``stiffness'' matrix~$A$~\cite{jin2015finite}, which is a function of $\varepsilon(\DOFsmooth(\vect{x}))$, so the subpixel smoothing occurs within the matrix assembly. The weak form is integrated element-by-element by a quadrature rule within each triangle, a weighted sum over quadrature points~\cite{jin2015finite}.  $\DOFfiltered$~is bilinearly interpolated from the Cartesian $\DOF$ grid, along with its gradient onto each quadrature point, at which point $\DOFsmooth$ (SSP) is computed and hence the material $\varepsilon$.  Because we are interpolating between air ($\Re \varepsilon > 0$) and metal ($\Re \varepsilon < 0$), it turns out to be preferable to compute $\varepsilon = n^2$ by linearly interpolating the complex refractive index $n$ with weight $\DOFsmooth$~\cite{testbed,rasmus_interp}.
The standard adjoint method for the operator yields the gradient of the objective function with respect to the materials at the quadrature points. It is then straightforward to further backpropagate through the projection, the bilinear interpolation~\cite{hammond20243}, and the filtering steps.

\begin{figure}
    \centering
    \includegraphics{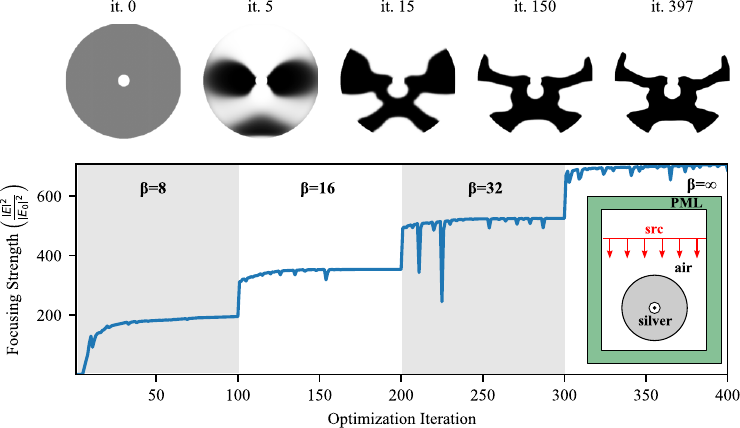}
    \caption{Topology optimization of a 2d metallic (Ag) particle to maximize field intensity $\Vert \vect{E} \Vert^2$ at the center of an annular design region (lower-right inset)~\cite{testbed}, using subpixel-smoothed projection and a finite-element method (FEM).  Figure of merit (focusing strength $\Vert \vect{E} \Vert^2$ relative to incident planewave amplitude $\Vert \vect{E}_0 \Vert^2$) vs.~iteration while increasing $\beta = 8,16,32,\infty$. Resulting FOM and design are similar to previous work~\cite{testbed}.
    }
    \label{fig:raman_TO}
\end{figure}

As in the previous sections, we performed density-based {\TopOpt}: we ran the optimization for 100 iterations at each $\beta$, for $\beta = 8.0, 16.0, 32.0, \infty$. \figref{fig:raman_TO} illustrates the geometric and the performance evolution.
Because surface-plasmon resonances in such nano-metallic structures can be very sensitive to small surface features, we evaluate the final objective value at high resolution, using a conforming contour of the mesh  constructed via the marching squares algorithm~\cite{maple2003geometric} applied to $\DOFprojected$ (not $\DOFsmooth$) with $\beta=\infty$. The results show a slight improvement in the objective function ($\approx 1130$) over the published solution ($\approx1030$)~\cite{christiansen2020inverse} in the testbed~\cite{testbed}, with a qualitatively similar geometry. This confirms that our smoothed projection is compatible with optimization over FEM discretizations on unstructured meshes, even for metallic/plasmonic structures where the fields are especially sensitive to interfaces.

\section{Conclusion}
\label{sec:conclusion}

\newtext{Although traditional density-{\TopOpt} projection schemes can be made to work by careful selection of $\beta$ schedules and/or penalty terms for intermediate materials---we and many other authors have used it successfully in many works!---it is attractive to eliminate such complications.}
We believe that the subpixel-smoothed projection (SSP) scheme introduced in this paper should be widely useful for topology optimization ({\TopOpt}) in many fields---nothing about the construction of our $\DOFsmooth$ is specific to electromagnetism.   SSP both simplifies and accelerates density-based {\TopOpt}, allowing $\beta$ to be increased very quickly without concern for ill-conditioning or ``frozen'' gradients, and at the same time guarantees almost-everywhere binarization of the final structure (without the need for additional penalty terms~\cite{van2013level,jensen2005damping, Hassan2022, hansen2024inverse, chen2025} in problems where the physics might otherwise push $\DOFfiltered$ to $\approx \eta$ in order to counteract finite-$\beta$ projection).  \newtext{That is, the shape can still be optimized at $\beta=\infty$, where SSP is constructed to guarantee that the geometry is binary except within $\sim \Delta x$ of interfaces (i.e.,~up to discretization), as illustrated in the inset of \figref{fig:diverging_optimization}.} Furthermore, because SSP simply replaces $\DOFprojected(\DOFfiltered)$ with a new function $\DOFsmooth(\DOFfiltered, \Vert \nabla \DOFfiltered \Vert)$, our approach should be easy to ``drop in'' to existing {\TopOpt} software without extensive modifications, and we have shown that it works well in both uniform-grid (e.g.~FDTD and FMM) and unstructured-mesh (FEM) discretizations.

Moreover, we believe that there are many opportunities for future developments building off of the smoothed-projection ideas presented in this paper.   We \newtext{are now preparing a manuscript that revisits} methods to impose fabrication-lengthscale constraints~\cite{zhou_geometric, Hammond21DRC}, which we believe can be further simplified and refined now that one can simply set $\beta = \infty$ to ensure binarization.  As mentioned in \secref{sec:topology_changes}, there is \newtext{also} the potential for a more sophisticated smoothed-projection scheme that allows one to differentiate through changes in topology for $\beta = \infty$ (instead of requiring $\beta$ to be increased in a few stages) by relaxing the $\DOFfiltered$-linearization approximation; \newtext{we are currently pursuing one promising approach}.  (Even with the current scheme, we often observe changes in topology at $\beta = \infty$ where the optimizer ``steps over'' the non-differentiability, e.g.~to delete a small hole, but this \newtext{may not be reliable} with algorithms whose convergence proofs assume differentiability.)  \newtext{It would also be useful to adapt existing schemes to automate the $\beta$ continuation~\cite{Dunning2025}, which is still useful to discover the correct topology and to nucleate holes/islands, to the rapid $\beta$ growth that is now practical.} Another exciting opportunity is to exploit the level-set information in $\DOFfiltered$ to improve \emph{accuracy} of the PDE discretization of the material interfaces---in other work, we've shown that one can improve accuracy in electromagnetism simply by using \emph{anisotropic smoothing}~\cite{FarjadpourRo06,OskooiKo09}, and there has also been other research on high-order discretization schemes exploiting subpixel interface knowledge~\cite{Law2022}.  In the finite-element context, we are working to both improve accuracy of the matrix assembly (quadrature over the mesh elements) and to allow smaller subpixel smoothing radii $\Rsmooth$, by adaptively subdividing the mesh in the vicinity of the level set, analogous to methods developed for immersed-boundary FEM techniques~\cite{Kumar2020} and level-set methods~\cite{van2013level}.

\appendix

\section*{Appendix A: Construction of smoothing kernels}

As explained in \secref{sec:smoothed_projection}, for $\beta \to \infty$ the subpixel-smoothed ``fill factor'' $F(d)$ has a particularly simple form \eqref{eq:F} that depends only on the distance $d$ from the current point $\vect{x}$ to the interface $\DOFprojected = \eta$ and on the smoothing kernel $\ksmooth(r)$.   Rather than specifying $\ksmooth(r)$ and computing $F(d)$, it is convenient to proceed in the reverse direction: construct a fill factor $F(d)$ with the desired properties (symmetry, differentiability, binary outside a compact domain, monotonicity) and infer $\ksmooth(r)$.   In this Appendix, we derive the straightforward algebraic relationship between $F(d)$ and $\ksmooth(r)$, and propose two possible choices of $F(d)$.

Differentiating \eqnref{eq:F} for $F(d)$ (with $d \ge 0$) with respect to~$d$ yields:
\begin{equation}
    F'(d) = -2\pi \int_0^{\pi/2} \ksmooth(d \sec \theta) d^2 \sec^2(\theta) \tan(\theta) \d \theta  = -2\pi \int_d^{\infty} \ksmooth(\xi) \xi \d \xi \,
\end{equation}
where we have let $\xi = d \sec \theta$, and hence taking a second derivative yields:
\begin{equation}
    F''(d) = 2\pi \ksmooth(d) d \, .
    \label{eq:k_from_F}
\end{equation}
Although this $F(d)$ was initially derived for $d \ge 0$ (i.e.~$\DOFprojected = 0$ at the origin), we can straightforwardly extend it to $d < 0$ by the mirror relation $F(-d) = 1 - F(d)$ (which follows from inverting $\DOFprojected \to 1 - \DOFprojected$ and then re-inverting the result).  This also implies that $F(0) = 1 - F(0) = \frac{1}{2}$.  We also impose the boundary condition $F(\infty) = 0$, since a localized smoothing kernel must yield zero when the interface is far away, along with $F(-\infty) = 1$ (when the interface is far in the other direction) and $F'(\pm\infty)=0$.
The unit-integral normalization of $\ksmooth(r)$ follows automatically from these relations, since we can integrate by parts to obtain:
\begin{align}
4\pi \int_0^{\infty} \ksmooth(r)r^2 \d r &= 2 \int_0^{\infty} F''(r) r \d r \\
&= \left. 2F'(r) r \right|_0^\infty - 2 \int_0^\infty F'(r) \d r \\
&= 2F(0) = 1 \, .
\end{align}
For a compactly supported $\ksmooth(r)$ that $=0$ for $r > \Rsmooth$, it also follows that $F(d) = 0$ for $d > \Rsmooth$ and $=1$ for $d < -\Rsmooth$.    To make $\DOFsmooth$ differentiable, it is necessary that $F(d)$ be differentiable, hence its derivative $F'$ must go to zero at $d = \pm \Rsmooth$.   Some optimization algorithms may implicitly \newtext{construct approximate second derivatives} (even if the \newtext{exact} second derivative is not explicitly supplied)~\cite{nocedal2006numerical,bertsekas2016optimization}, \newtext{or assume twice differentiability in their convergence proofs~\cite{Svanberg2002,Wchter2005},} so it is also desirable to have $F''$ be continuous (i.e.,~$F''$ should go to zero at $d = \pm \Rsmooth$). 

The computationally cheapest twice-differentiable $F(d)$ is probably a piecewise polynomial.  The symmetry relation implies that $F(d)-\frac{1}{2}$ must be odd (containing only odd powers of $d$).  The six boundary conditions (on $F, F', F''$) at $d=\pm \Rsmooth$ reduce to three conditions for an odd function, and hence we must have at least three free coefficients.  This means that $F(d)$ must be at least a degree-5 polynomial, and solving algebraically for the coefficients satisfying the boundary conditions leads to the polynomial function $F_2$ of \eqnref{eq:Fpoly}, which also turns out to be monotonic.
By \eqnref{eq:k_from_F}, the corresponding smoothing kernel is a continuous, concave, quadratic ``bump'':
\begin{equation}
    \ksmooth_2(r) = \frac{F_2''(r)}{2\pi r} = \frac{1}{2\pi \Rsmooth^3}\begin{cases}
        \frac{15}{4} \left[ 1 - \left( \frac{r}{\Rsmooth} \right)^2 \right] & r \le \Rsmooth \\
        0 & r > \Rsmooth
    \end{cases} \, , 
    \label{eq:kpoly}
\end{equation}
although we do not actually use this $\ksmooth(r)$ directly for any calculation.

If one has a high-order PDE discretization or an optimization algorithm that relies on the existence of third or higher derivatives, it might be desirable to have an even smoother $F(d)$.   Rather than going to ever higher degrees of polynomials, it may be convenient for such cases to simply construct an \emph{infinitely} differentiable ($C^\infty$) fill factor. For this purpose, we could employ, for example, a well-known $C^\infty$ monotonic ``transition function''~\cite[ch.~2]{nestruev2010}:
\begin{equation}
    \tau(u) = \begin{cases}
        \frac{1}{1 + \exp\left(\frac{1}{u}-\frac{1}{1-u}\right)} & 0 < u < 1 \\
        0 & u \le 0 \\
        1 & u \ge 1
    \end{cases} \, ,
\end{equation}
which goes smoothly from $\tau(0)=0$ to $\tau(1)=1$ (all derivatives vanish at $u=0^+$ and $u=1^-$), satisfying $\tau(1-u)=1-\tau(u)$.
In terms of $\tau(u)$, we can then define: 
\begin{equation}
F_\infty(d) = \tau\left(\frac{1}{2} - \frac{d}{2\Rsmooth}\right) \,
\label{eq:Fsmooth}
\end{equation}
for which all derivatives go to zero as $d \to \pm \Rsmooth$, and it satisfies all of our other conditions on~$F(d)$.  (Of course, there are many other choices of $C^\infty$ functions. This particular $F_\infty$ has about twice the slope of our polynomial $F_2$ at $d=0$, so to get a similar-looking $\DOFsmooth$ one would need to roughly double the smoothing radius~$\Rsmooth$.)

\section*{Appendix B: Resolution-independent second-derivatives}

As discussed in the main text, traditional {\TopOpt} employing~$\DOFprojected$ tends to converge slowly for large finite~$\beta$ because of the large second derivatives of the projection function~$P_{\beta,\eta}$, which lead to an ill-conditioned Hessian (second-derivative) matrix for any objective function depending on~$\DOFprojected$.   At first glance, it may seem that our new~$\DOFsmooth$ has the same problem at high resolutions (small~$\Delta x$), since by construction we have~$\DOFsmooth \to \DOFprojected$ as~$\Delta x \to 0$.  Fortunately, this does \emph{not} lead to ill-conditioning at high resolutions: the derivatives of~$\DOFsmooth$ with respect to \emph{space}~($\vect{x}$) diverge as~$\Delta x \to 0$ for~$\beta = \infty$, but \emph{not} the derivatives with respect to the discretized~$\DOFfiltered$ values.

This is easiest to see with a 1d example. Suppose that we have~$\DOFfiltered$ on an equispaced 1d finite-difference grid~$\DOFfiltered_k = \DOFfiltered(k\Delta x)$, and we wish to evaluate~$\DOFsmooth$ at a point~$(n+\frac{1}{2})\Delta x$ halfway in between two grid points~$n$ and~$n+1$.  To evaluate~$d$ from \eqnref{eq:d}, we might perform a piecewise-linear interpolation of~$\DOFfiltered$, giving
$$
d = \Delta x \frac{\eta - \frac{\DOFfiltered_n + \DOFfiltered_{n+1}}{2}}{|\DOFfiltered_{n+1}-\DOFfiltered_n|} \, ,
$$
which is~$\Delta x$ multiplied by a dimensionless function of~$\DOFfiltered$.  However, when we plug this into~$F(d)$ to evaluate~$\DOFsmooth$, the~$\Delta x$ factor \emph{cancels}: $F(d)$ is a function of~$d/\Rsmooth$, and~$\Rsmooth$ is proportional to~$\Delta x$.   Hence, the derivatives of~$\DOFsmooth$ at this grid point \emph{with respect to}~$\DOFfiltered_k$ are bounded regardless of resolution.  (This is true even when the denominator of $d$ approaches zero, since in that regime $F$ is constant so all derivatives vanish.)

The same fortunate effect also holds for more general meshes/discretizations, as long as $\DOFfiltered(\vect{x})$ is represented in terms of values sampled at a set of points in space (e.g.~FEM nodes), because dimensional considerations ensure that~$\nabla \DOFfiltered$ must then be~$\frac{1}{\Delta x}$ multiplied by a dimensionless ($\Delta x$-independent) linear combination of the~$\DOFfiltered$ samples.  Hence,~$d/\Rsmooth$ will always be a~$\Delta x$-independent function of the discretized~$\DOFfiltered$ values, and a similar cancellation occurs for the~$\Rsmooth \Vert \nabla \DOFfiltered \Vert$ factor in \eqnref{eq:rhohat_pm}.

\section*{Appendix C: Example Python implementation}
\newtext{The following is a differentiable Python implementation of SSP on a 2d Cartesian grid using the Autograd
package~\cite{maclaurin2015autograd}.   Below, the function \code{smoothed\_projection} takes as input a 2d array \code{rho\_filt} representing $\DOFfiltered$, the $\beta$ and $\eta$ parameters of the projection, and the \code{resolution} $1/\Delta x$ of the grid, returning the smoothed projection $\DOFsmooth$.  (Here, $\nabla \DOFfiltered$ is computed by centered differences using the \code{autograd.numpy.gradient} function; the code could easily be modified to take $\Vert \nabla \DOFfiltered \Vert^2$ as a parameter instead.  Some care is taken, below, to avoid floating-point exceptions in Autograd when $\Vert \nabla \DOFfiltered \Vert = 0$.)  This code can also be found in the Meep FDTD software~\cite{meep}.}
{\footnotesize
\input{minted-cache/listing1.pygtex}
}

\begin{backmatter}
\bmsection{Funding}
This material is based upon work supported in part by the National Science Foundation (NSF) Center ``EPICA'' under Grant No.1 2052808, \url{https://epica.research.gatech.edu/}. Any opinions, findings, and conclusions or recommendations expressed in this material are those of the author(s) and do not necessarily reflect the  views of the NSF. MC, IMH, and SGJ were supported in part by the US Army Research Office through the Institute for Soldier Nanotechnologies (award W911NF-23-2-0121) and by a grant from the Simons Foundation.  SER was supported by the Georgia Electronic Design Center of the Georgia Institute of Technology.

\bmsection{Acknowledgments}
We are grateful to Rodrigo Arrieta Candia and Giuseppe Romano at MIT for helpful comments, and to Francesc Verdugo at Vrije Universiteit Amsterdam for assistance with the Gridap FEM software.

\bmsection{Disclosures}
The authors declare no conflicts of interest.

\bmsection{Data availability} Data underlying the results presented in this paper are available upon request.

\end{backmatter}


\bibliography{refs}

\begin{thebibliography}{10}
\newcommand{\enquote}[1]{``#1''}

\bibitem{Molesky2018}
S.~Molesky, Z.~Lin, A.~Y. Piggott, W.~Jin, J.~Vuckovi{\'{c}}, and A.~W.
  Rodriguez, \enquote{Inverse design in nanophotonics,}
  {\protect\JournalTitle{Nature Photonics}} \textbf{12}, 659--670 (2018).

\bibitem{van2013level}
N.~P. van Dijk, K.~Maute, M.~Langelaar, and F.~Van~Keulen, \enquote{Level-set
  methods for structural topology optimization: A review,}
  {\protect\JournalTitle{Structural and Multidisciplinary Optimization}}
  \textbf{48}, 437--472 (2013).

\bibitem{to_approaches}
O.~Sigmund and K.~Maute, \enquote{Topology optimization approaches,}
  {\protect\JournalTitle{Structural and Multidisciplinary Optimization}}
  \textbf{48}, 1031--1055 (2013).

\bibitem{Jensen2010}
J.~Jensen and O.~Sigmund, \enquote{Topology optimization for nano‐photonics,}
  {\protect\JournalTitle{Laser \& Photonics Reviews}} \textbf{5}, 308–321
  (2010).

\bibitem{nocedal2006numerical}
J.~Nocedal and S.~J. Wright, \emph{Numerical Optimization} (Springer, 2006),
  2nd ed.

\bibitem{bertsekas2016optimization}
D.~P. Bertsekas, \emph{Nonlinear Programming} (Athena Scientific, 2016), 3rd
  ed.

\bibitem{meep_adjoint}
A.~M. Hammond, A.~Oskooi, M.~Chen, Z.~Lin, S.~G. Johnson, and S.~E. Ralph,
  \enquote{High-performance hybrid time/frequency-domain topology optimization
  for large-scale photonics inverse design,} {\protect\JournalTitle{Optics
  Express}} \textbf{30}, 4467--4491 (2022).

\bibitem{projection_overview}
B.~S. Lazarov, F.~Wang, and O.~Sigmund, \enquote{Length scale and
  manufacturability in density-based topology optimization,}
  {\protect\JournalTitle{Archive of Applied Mechanics}} \textbf{86}, 189--218
  (2016).

\bibitem{linus_morph}
L.~H{\"a}gg and E.~Wadbro, \enquote{On minimum length scale control in density
  based topology optimization,} {\protect\JournalTitle{Structural and
  Multidisciplinary Optimization}} \textbf{58}, 1015--1032 (2018).

\bibitem{qian_actuators}
X.~Qian and O.~Sigmund, \enquote{Topological design of electromechanical
  actuators with robustness toward over-and under-etching,}
  {\protect\JournalTitle{Computer Methods in Applied Mechanics and
  Engineering}} \textbf{253}, 237--251 (2013).

\bibitem{zhou_geometric}
M.~Zhou, B.~S. Lazarov, F.~Wang, and O.~Sigmund, \enquote{Minimum length scale
  in topology optimization by geometric constraints,}
  {\protect\JournalTitle{Computer Methods in Applied Mechanics and
  Engineering}} \textbf{293}, 266--282 (2015).

\bibitem{svanberg_harmonic}
K.~Svanberg and H.~Sv{\"a}rd, \enquote{Density filters for topology
  optimization based on the {Pythagorean} means,}
  {\protect\JournalTitle{Structural and Multidisciplinary Optimization}}
  \textbf{48}, 859--875 (2013).

\bibitem{Guest2011}
J.~K. Guest, A.~Asadpoure, and S.-H. Ha, \enquote{Eliminating beta-continuation
  from heaviside projection and density filter algorithms,}
  {\protect\JournalTitle{Structural and Multidisciplinary Optimization}}
  \textbf{44}, 443–453 (2011).

\bibitem{Dunning2025}
P.~Dunning and F.~Wein, \enquote{Automatic projection parameter increase for
  three-field density-based topology optimization,}
  {\protect\JournalTitle{Structural and Multidisciplinary Optimization}}
  \textbf{68} (2025).

\bibitem{jensen2005damping}
J.~S. Jensen and O.~Sigmund, \enquote{Topology optimization of photonic crystal
  structures: A high-bandwidth low-loss {T}-junction waveguide,}
  {\protect\JournalTitle{JOSA B}} \textbf{22}, 1191--1198 (2005).

\bibitem{hansen2024inverse}
S.~E. Hansen, G.~Arregui, A.~N. Babar, R.~E. Christiansen, and S.~Stobbe,
  \enquote{Inverse design and characterization of compact, broadband, and
  low-loss chip-scale photonic power splitters,}
  {\protect\JournalTitle{Materials for Quantum Technology}} \textbf{4}, 016201
  (2024).

\bibitem{chen2025}
M.~Chen, K.~F. Chan, A.~M. Hammond, C.~H. Chan, and S.~G. Johnson,
  \enquote{Inverse design of 3d-printable metalenses with complementary
  dispersion for terahertz imaging,}  (2025). ArXiv:2502.10520.

\bibitem{Rudin_1976}
W.~Rudin, \emph{Principles of Mathematical Analysis} (McGraw-Hill, 1976), 3rd
  ed.

\bibitem{taflove_em_book}
A.~Taflove and S.~C. Hagness, \emph{Computational Electrodynamics: The
  Finite-Difference Time-Domain Method} (Artech house, 2005).

\bibitem{kim2012fourier}
H.~Kim, J.~Park, and B.~Lee, \emph{Fourier Modal Method and Its Applications in
  Computational Nanophotonics} (CRC Press Boca Raton, 2012).

\bibitem{jin2015finite}
J.-M. Jin, \emph{The Finite Element Method in Electromagnetics} (John Wiley \&
  Sons, 2015).

\bibitem{testbed}
M.~Chen, R.~E. Christiansen, J.~A. Fan, G.~I{\c{s}}iklar, J.~Jiang, S.~G.
  Johnson, W.~Ma, O.~D. Miller, A.~Oskooi, M.~F. Schubert \emph{et~al.},
  \enquote{Validation and characterization of algorithms and software for
  photonics inverse design,} {\protect\JournalTitle{Journal of the Optical
  Society of America B}} \textbf{41}, A161--A176 (2024).

\bibitem{deRuiter2004}
M.~de~Ruiter and F.~van Keulen, \enquote{Topology optimization using a topology
  description function,} {\protect\JournalTitle{Structural and
  Multidisciplinary Optimization}} \textbf{26}, 406–416 (2004).

\bibitem{van2012explicit}
N.~Van~Dijk, M.~Langelaar, and F.~Van~Keulen, \enquote{Explicit level-set-based
  topology optimization using an exact heaviside function and consistent
  sensitivity analysis,} {\protect\JournalTitle{International Journal for
  Numerical Methods in Engineering}} \textbf{91}, 67--97 (2012).

\bibitem{rezk2005vertex}
C.~Rezk-Salama and A.~Kolb, \enquote{A vertex program for efficient box-plane
  intersection,} in \emph{Proc. Vision, Modeling and Visualization,}  (2005),
  pp. 115--122.

\bibitem{Wein2020}
F.~Wein, P.~D. Dunning, and J.~A. Norato, \enquote{A review on feature-mapping
  methods for structural optimization,} {\protect\JournalTitle{Structural and
  Multidisciplinary Optimization}} \textbf{62}, 1597–1638 (2020).

\bibitem{Svanberg2002}
K.~Svanberg, \enquote{A class of globally convergent optimization methods based
  on conservative convex separable approximations,} {\protect\JournalTitle{SIAM
  journal on optimization}} \textbf{12}, 555--573 (2002).

\bibitem{Hammond21DRC}
A.~Hammond, A.~Oskooi, S.~Johnson, and S.~Ralph, \enquote{Photonic topology
  optimization with semiconductor-foundry design-rule constraints,}
  {\protect\JournalTitle{Optics Express}} \textbf{29} (2021).

\bibitem{White2018}
D.~A. White, M.~L. Stowell, and D.~A. Tortorelli, \enquote{Toplogical
  optimization of structures using {Fourier} representations,}
  {\protect\JournalTitle{Structural and Multidisciplinary Optimization}}
  \textbf{58}, 1205–1220 (2018).

\bibitem{Sanders2021}
C.~Sanders, M.~Bonnet, and W.~Aquino, \enquote{An adaptive eigenfunction basis
  strategy to reduce design dimension in topology optimization,}
  {\protect\JournalTitle{International Journal for Numerical Methods in
  Engineering}} \textbf{122}, 7452–7481 (2021).

\bibitem{Chandrasekhar2022}
A.~Chandrasekhar and K.~Suresh, \enquote{Approximate length scale filter in
  topology optimization using {Fourier} enhanced neural networks,}
  {\protect\JournalTitle{Computer-Aided Design}} \textbf{150}, 103277 (2022).

\bibitem{rasmus_interp}
R.~E. Christiansen, J.~Vester-Petersen, S.~P. Madsen, and O.~Sigmund,
  \enquote{A non-linear material interpolation for design of metallic
  nano-particles using topology optimization,} {\protect\JournalTitle{Computer
  Methods in Applied Mechanics and Engineering}} \textbf{343}, 23--39 (2019).

\bibitem{Hassan2017}
E.~Hassan, E.~Wadbro, L.~H\"{a}gg, and M.~Berggren, \enquote{Topology
  optimization of compact wideband coaxial-to-waveguide transitions with
  minimum-size control,} {\protect\JournalTitle{Structural and
  Multidisciplinary Optimization}} \textbf{57}, 1765–1777 (2018).

\bibitem{Schevenels2016morph}
M.~Schevenels and O.~Sigmund, \enquote{On the implementation and effectiveness
  of morphological close-open and open-close filters for topology
  optimization,} {\protect\JournalTitle{Structural and Multidisciplinary
  Optimization}} \textbf{54}, 15–21 (2016).

\bibitem{Wang2011}
F.~Wang, B.~S. Lazarov, and O.~Sigmund, \enquote{On projection methods,
  convergence and robust formulations in topology optimization,}
  {\protect\JournalTitle{Structural and Multidisciplinary Optimization}}
  \textbf{43}, 767--784 (2011).

\bibitem{meep}
A.~F. Oskooi, D.~Roundy, M.~Ibanescu, P.~Bermel, J.~D. Joannopoulos, and S.~G.
  Johnson, \enquote{{MEEP}: A flexible free-software package for
  electromagnetic simulations by the {FDTD} method,}
  {\protect\JournalTitle{Computer Physics Communications}} \textbf{181},
  687--702 (2010).

\bibitem{maclaurin2015autograd}
D.~Maclaurin, D.~Duvenaud, and R.~P. Adams, \enquote{Autograd: Effortless
  gradients in numpy,} in \emph{ICML 2015 AutoML Workshop,}  vol. 238 (2015),
  p.~5.

\bibitem{wu2020state}
S.~Wu, X.~Mu, L.~Cheng, S.~Mao, and H.~Fu, \enquote{State-of-the-art and
  perspectives on silicon waveguide crossings: A review,}
  {\protect\JournalTitle{Micromachines}} \textbf{11}, 326 (2020).

\bibitem{fmmax}
M.~F. Schubert and A.~M. Hammond, \enquote{Fourier modal method for inverse
  design of metasurface-enhanced micro-{LED}s,} {\protect\JournalTitle{Optics
  Express}} \textbf{31}, 42945--42960 (2023).

\bibitem{Badia2020}
S.~Badia and F.~Verdugo, \enquote{Gridap: An extensible finite element toolbox
  in {Julia},} {\protect\JournalTitle{Journal of Open Source Software}}
  \textbf{5}, 2520 (2020).

\bibitem{snyder_love}
A.~W. Snyder and J.~Love, \emph{Optical Waveguide Theory} (Springer Science \&
  Business Media, 2012).

\bibitem{JohnsonMa98}
S.~G. Johnson, C.~Manolatou, S.~Fan, P.~Villeneuve, J.~D. Joannopoulos, and
  H.~A. Haus, \enquote{Elimination of cross talk in waveguide intersections,}
  {\protect\JournalTitle{Optics Letters}} \textbf{23}, 1855--1857 (1998).

\bibitem{nlopt}
S.~G. Johnson, \enquote{The {NLopt} nonlinear-optimization package,}
  \url{https://github.com/stevengj/nlopt} (2007).

\bibitem{adam}
D.~P. Kingma and J.~Ba, \enquote{Adam: A method for stochastic optimization,}
  (2014).

\bibitem{sell2017large}
D.~Sell, J.~Yang, S.~Doshay, R.~Yang, and J.~A. Fan, \enquote{Large-angle,
  multifunctional metagratings based on freeform multimode geometries,}
  {\protect\JournalTitle{Nano Letters}} \textbf{17}, 3752--3757 (2017).

\bibitem{sell2018ultra}
D.~Sell, J.~Yang, E.~W. Wang, T.~Phan, S.~Doshay, and J.~A. Fan,
  \enquote{Ultra-high-efficiency anomalous refraction with dielectric
  metasurfaces,} {\protect\JournalTitle{ACS Photonics}} \textbf{5}, 2402--2407
  (2018).

\bibitem{sell2017periodic}
D.~Sell, J.~Yang, S.~Doshay, and J.~A. Fan, \enquote{Periodic dielectric
  metasurfaces with high-efficiency, multiwavelength functionalities,}
  {\protect\JournalTitle{Advanced Optical Materials}} \textbf{5}, 1700645
  (2017).

\bibitem{jiang2019global}
J.~Jiang and J.~A. Fan, \enquote{Global optimization of dielectric metasurfaces
  using a physics-driven neural network,} {\protect\JournalTitle{Nano Letters}}
  \textbf{19}, 5366--5372 (2019).

\bibitem{jiang2020simulator}
J.~Jiang and J.~A. Fan, \enquote{Simulator-based training of generative neural
  networks for the inverse design of metasurfaces,}
  {\protect\JournalTitle{Nanophotonics}} \textbf{9}, 1059--1069 (2020).

\bibitem{schubert2024invrsgymtoolkitnanophotonicinverse}
M.~F. Schubert, \enquote{invrs-gym: a toolkit for nanophotonic inverse design
  research,}  (2024).

\bibitem{jax}
J.~Bradbury, R.~Frostig, P.~Hawkins, M.~J. Johnson, C.~Leary, D.~Maclaurin,
  G.~Necula, A.~Paszke, J.~Vander{P}las, S.~Wanderman-{M}ilne, and Q.~Zhang,
  \enquote{{JAX}: composable transformations of {P}ython+{N}um{P}y programs,}
  \url{http://github.com/google/jax} (2018).

\bibitem{christiansen2020inverse}
R.~E. Christiansen, J.~Michon, M.~Benzaouia, O.~Sigmund, and S.~G. Johnson,
  \enquote{Inverse design of nanoparticles for enhanced raman scattering,}
  {\protect\JournalTitle{Optics Express}} \textbf{28}, 4444--4462 (2020).

\bibitem{hammond20243}
I.~M. Hammond, \enquote{3-{D} topology optimization of spatially averaged
  surface-enhanced {Raman} devices,} Master's thesis, Massachusetts Institute
  of Technology (2024).

\bibitem{maple2003geometric}
C.~Maple, \enquote{Geometric design and space planning using the marching
  squares and marching cube algorithms,} in \emph{Proc. 2003 Intl. Conf. on
  Geometric Modeling and Graphics,}  (IEEE, 2003), pp. 90--95.

\bibitem{Hassan2022}
E.~Hassan and A.~Cal{\`a}~Lesina, \enquote{Topology optimization of dispersive
  plasmonic nanostructures in the time-domain,} {\protect\JournalTitle{Optics
  Express}} \textbf{30}, 19557 (2022).

\bibitem{FarjadpourRo06}
A.~Farjadpour, D.~Roundy, A.~Rodriguez, M.~Ibanescu, P.~Bermel, J.~D.
  Joannopoulos, S.~G. Johnson, and G.~Burr, \enquote{Improving accuracy by
  subpixel smoothing in {FDTD},} {\protect\JournalTitle{Optics Letters}}
  \textbf{31}, 2972--2974 (2006).

\bibitem{OskooiKo09}
A.~F. Oskooi, C.~Kottke, and S.~G. Johnson, \enquote{Accurate finite-difference
  time-domain simulation of anisotropic media by subpixel smoothing,}
  {\protect\JournalTitle{Optics Letters}} \textbf{34}, 2778--2780 (2009).

\bibitem{Law2022}
Y.-M. Law and J.-C. Nave, \enquote{High-order {FDTD} schemes for {Maxwell}’s
  interface problems with discontinuous coefficients and complex interfaces
  based on the correction function method,} {\protect\JournalTitle{Journal of
  Scientific Computing}} \textbf{91} (2022).

\bibitem{Kumar2020}
A.~V. Kumar, \enquote{Survey of immersed boundary approaches for finite element
  analysis,} {\protect\JournalTitle{Journal of Computing and Information
  Science in Engineering}} \textbf{20} (2020).

\bibitem{Wchter2005}
A.~W\"{a}chter and L.~T. Biegler, \enquote{On the implementation of an
  interior-point filter line-search algorithm for large-scale nonlinear
  programming,} {\protect\JournalTitle{Mathematical Programming}} \textbf{106},
  25–57 (2005).

\bibitem{nestruev2010}
J.~Nestruev, \emph{Smooth Manifolds and Observables} (Springer, 2010).

\end{thebibliography}

\end{document}